\documentstyle[art10]{article}
\twocolumn
\renewcommand{\title}[1]{\gdef\titlex{#1}}
\renewcommand{\author}[1]{\gdef\authorx{#1}}
\newcommand{\institute}[1]{\gdef\institutex{#1}}
\newcommand{\titlerunning}[1]{ }
\newcommand{\strich}{\noindent\hrulefill \ }
\newcommand{\E}{\mbox{\rm e}}
\renewcommand{\maketitle}{\twocolumn[{\newpage\vspace*{1cm}\noindent
  {\LARGE\bf \titlex}\vspace{12pt}\newline {\bf \authorx}
  \vspace{7pt}\\ {\institutex}\vspace{7pt}
  \\ {To be published in Z.Physik B. Submitted Aug. 1995.
  Revised Oct. 5. 1995.}\vspace{12pt}\\ }]}
\renewcommand{\theequation}{\arabic{section}.\arabic{equation}}
\newcommand{\bm}[1]{\mbox{\boldmath$#1$}}
\newcommand{\sfrac}[2]{{\scriptstyle \frac{#1}{#2}}}
\newcommand{\av}[1]{\big\langle #1 \big\rangle}
\renewcommand{\d}[1]{\mbox{d}#1\,}
\newcommand{\otop}{\mathaccent"7617}
\newcommand{\toto}{\mathop{\longrightarrow}}
\newcommand{\citeq}[1]{(\ref{#1})}
\newcommand{\mfigure}[3]{\begin{figure}[ht] \epsfxsize=223pt
  \epsfbox{#1.eps} \caption{#2} \label{#3}\end{figure}}
\newenvironment{acknowledgement}{\vspace{0.3cm}\footnotesize\noindent
  {\em Acknowledgement.}}{}
\renewenvironment{abstract}{{\noindent\bf Abstract. }\noindent}{}

\input epsf
\epsfverbosetrue

\begin{document}

\title{Drift, creep and pinning of a particle in a correlated
       \\ random potential}
\titlerunning{Drift, creep and pinning}
\author{Heinz Horner}
\institute{Institut f\"ur Theoretische Physik, Universit\"at Heidelberg,
Philosophenweg 19, D-69120 Heidelberg}
\maketitle

\begin{abstract} The motion of a particle in a correlated random potential
under the influence of a driving force is investigated in mean field theory.
The correlations of the disorder are characterized by a short distance cutoff
and a power law decay with exponent $\gamma$ at large distances. Depending on
temperature and $\gamma$ , drift with finite mobility, creep or pinning is
found. This is in qualitative agreement with results in one dimension. This
model is of interest not only in view of the motion of particles or manifolds
in random media, it also improves the understanding of glassy non-equilibrium
dynamics in mean field models. The results, obtained by numerical integration
and analytic investigations of the various scaling regimes in this problem, are
compared with previous proposals regarding the long time properties of such
systems and with replica calculations.
\end{abstract}

\strich

\section{Introduction}
\setcounter{equation}{0}

This paper deals with the motion of a particle in a correlated random
potential under the influence of a driving force. The correlations of the
disorder are characterized by a short distance cutoff and a power law decay at
larger distances with exponent $\gamma$. This problem has been solved in one
dimension
\cite{Sinai82,leDou95,Scheidl95} and, depending on $\gamma$, drift with finite
mobility, creep or pinning was found. Creep means that the particle moves with
a mean velocity less than proportional to the driving force and in case of
pinning the mean velocity is zero, unless the force exceeds some critical
value.

In $N$ dimensions a mean field treatment becomes exact for $N\to\infty$ and
there is a close formal similarity to spin glass problems with long range
interactions. The situation without external driving force has been treated
within replica theory \cite{Mez:manifold90,Mez:manifold91} and also using a
stochastic dynamics approach \cite{KibachHo93I,KibachHo93II}. Actually the
more general case of the motion of a $D$-dimensional manifold in an
$N\to\infty$ dimensional space was investigated. For finite $N$ the mean field
treatment is only an approximation. Nevertheless it might give some clue about
a great variety of systems with disorder, for instance flux lines in
superconductors or interfaces in random field systems.

The close formal relation to spin glass problems raises another set of
questions. There has been a continuous interest in the dynamics of spin
glasses and related systems. The pioneering work of Sompolinsky and Zippelius
\cite{Somp81,SZ82} concentrated in part on an understanding of the replica
theory and the hierarchical replica symmetry breaking scheme proposed by
Parisi \cite{Par79}. The basic assumption was about the existence of a
hierarchy of diverging time scales with ultrametric properties. This point has
been investigated further \cite{Ho:DySK84} introducing a long time scale $\bar
t$ associated with slow changes of the random interactions in this model. It
could be shown that the above assumption meant that the long time
contributions to the correlation and response functions have to be smooth
functions of $x(t)=1-\ln t/\ln\bar t$ with $\bar t\to\infty$.
It was later shown \cite{Ho:DySK87} that this led to inconsistencies
which could not be resolved at that time.

There exists a variety of models with disorder where a replica treatment
requires only single step replica symmetry breaking instead of the
hierarchical scheme mentioned above. The resulting phase transition can be
considered as discontinuous in contrast to the continuous transition observed
in the SK-model and other cases requiring hierarchical replica symmetry
breaking. A treatment via dynamics
\cite{KibachHo93I,KiTh87,ho:binperc92,CHS93} revealed a dynamic freezing
temperature above the transition temperature obtained in replica theory. This
is in contrast to the continuous transitions where the same transition
temperature is found. Furthermore it could be shown \cite{CHS93} that the
states which contribute to dynamics have a higher energy than the states
relevant for the replica calculation. Numerical simulations of a learning
process in a perceptron with binary bonds
\cite{ho:binperc92}, another system with discontinuous transition, indicated
that the replica result applies if the limit $\bar t\to\infty$ is performed
first and then the limit $N\to\infty$, whereas the results obtained via
dynamics hold for the opposite order. This observation suggests that in the
thermodynamic limit the particular states most relevant for the replica
calculation cannot be reached within any finite time. This picture is
consistent with the fact that finding the best groundstate in a spin glass or
perfect learning in a perceptron with binary bonds is a combinatorial hard
problem which cannot be solved in polynomial time. The replica treatment, on
the other hand, is purely static and does not refer to any kind of dynamics.
For a given physical situation the appropriate order of limits has of course
to be found, keeping in mind that the longest equilibration times are likely
to diverge exponentially with some power of $N$ \cite{KibachHo91}.

This raises the question whether similar discrepancies also show up in systems
with continuous transition. The problem of the motion of a particle in a
random potential under the influence of a driving force is in several aspects
an appropriate model system to address this question. Prescribing the drift
velocity $v$ and calculating the force necessary to drive the particle, a long
external time scale is defined by $\bar t\sim v^{-1}$. For times exceeding
$\bar t$ the system is supposed to reach a stationary non-equilibrium state, a
so-called flow state, where correlation and response functions depend on
differences of time only. This is of great advantage regarding numerical as
well as analytic work. Depending on the exponent $\gamma$ and on temperature,
a continuous as well as a discontinuous transition can be found and therefore
both cases can be studied on a single model.

The results presented in the following are based on numerical integrations of
the dynamic mean field equations of this model, extending over more than
thirty orders of magnitude in time and velocity. Such a wide range is actually
necessary in order to understand the different scaling and crossover regimes
emerging in this problem. Such a wide range of time scales is not unrealistic
keeping in mind that phenomena might be observed on a scale of days or more
with an intrinsic time scale of $10^{-12}$ to
$10^{-16}$ seconds typical for vibrational or electronic degrees of freedom.

The numerical results are supplemented by exhaustive analytic work
characterizing the different scaling regimes and focusing on their asymptotic
properties and associated exponents. The characteristic time scales are
obtained by matching between adjacent scaling regimes and are again governed
by characteristic exponents.

The key result of the present investigation is the observation that the
dynamics with a long but finite external time scale $\bar t$ is ruled by three
regimes:
\begin{description}
\item[i)]   The FDT-regime describing local equilibrium at short times.
\item[ii)]  The intermediate plateau regime for times $t\sim {\bar t}^{\,\xi}$
with $0<\xi<1$. 
\item[iii)] The asymptotic regime for $t\sim \bar t$.
\end{description} These regimes have additional substructures and associated
additional internal time scales.

The plateau regime is the crucial link between short times and long times of
the order of the external time scale. Its properties determine for instance
whether creep or pinning is found or whether the longest internal time scale
is proportional to $\bar t$ or $\bar t^{\,1+\eta}$ with $\eta>0$. This
question is of relevance for instance in the context of aging phenomena
\cite{aging1,aging2,aging3}.

\mfigure{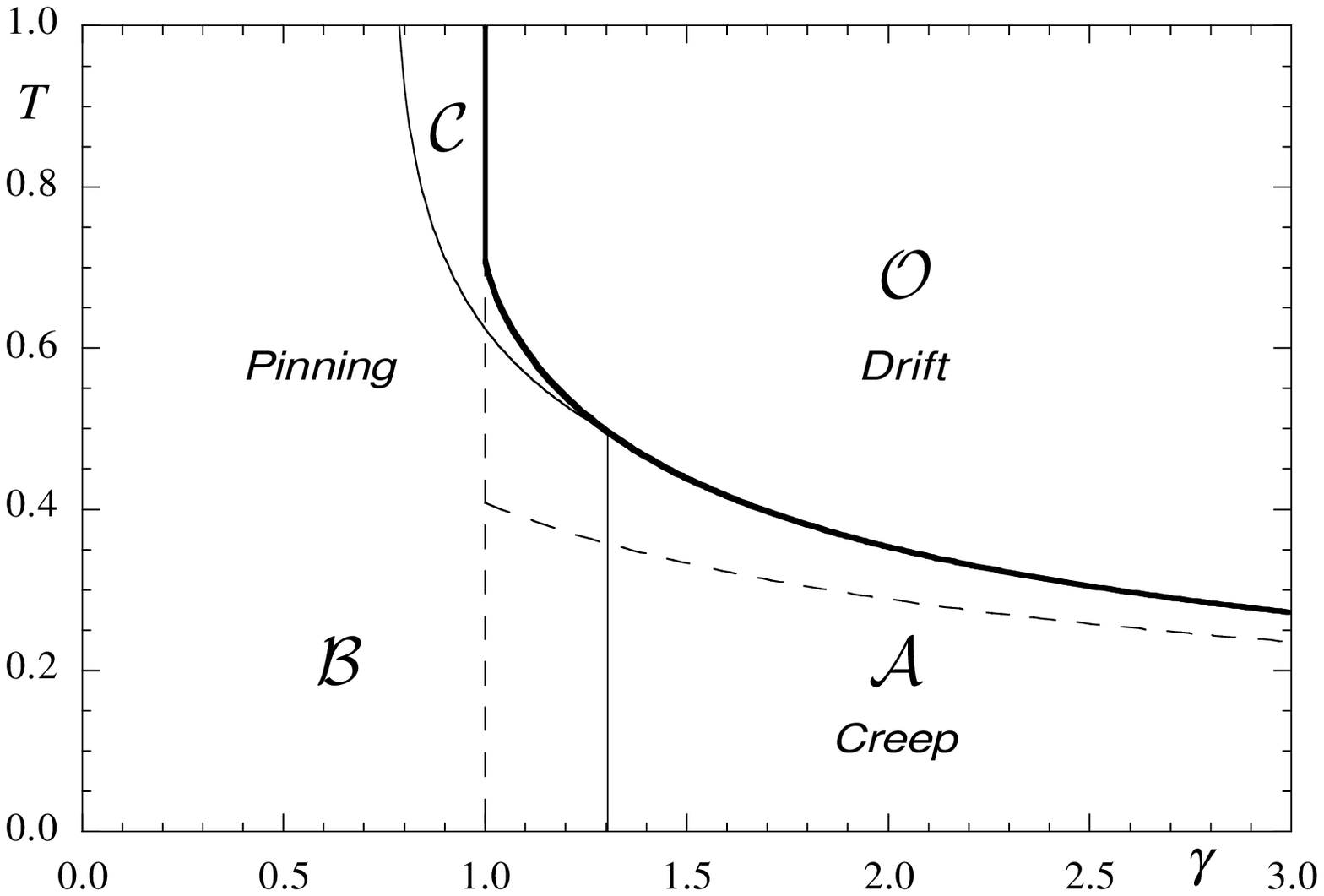} {Phase diagram. $\cal O$: Drift phase with $v\sim F$
(driving force $F$, resulting velocity $v$). $\cal A$: Creep phase with $v\sim
F^{1/\eta}$. $\cal B$ and
$\cal C$: Pinning phases with $v=0$ for $F<F_p$. The phase diagram resulting
from replica theory is also shown (doted lines)}{phase-fig}

Depending on $\gamma$ and temperature, different phases are found. The phase
diagram resulting from the present investigations is shown in
Fig.\ref{phase-fig}. There is a phase with finite mobility ($\cal O$), another
one with creep behavior ($\cal A$) and two more where pinning is observed
($\cal B$ and $\cal C$). This is in qualitative agreement with the findings
for the one-dimensional model \cite{leDou95}, which indicates that the
transitions are not an artifact of the mean field treatment. In all phases,
except the high temperature phase $\cal O$, additional intermediate time
scales exist, which diverge for $v\to0$. There is, however, no ultrametric
organization of time scales. This is in contrast to the common proposal
\cite{KibachHo93II,SZ82,Ho:DySK84,CHS93}. Moreover, the phase diagram
suggested by replica theory \cite{Mez:manifold90,Mez:manifold91} or dynamics
within the hypothesis of ultrametric time scales
\cite{KibachHo93I,KibachHo93II} differs from the one obtained in the present
investigation. For instance, $\gamma=1$ has been predicted for the $\cal
A$-$\cal B$ phase boundary in contrast to $\gamma=1.3044$ found at present.
This leads to the conclusion that the states dominating the replica theory
with broken symmetry are again not the same as those relevant for dynamics on
long but finite time scales.

As a consequence of the assumption of ultrametricity in the organization of
diverging time scales, the resulting correlation and response functions were
not unique
\cite{SZ82,Ho:DySK84}. The present investigation yields on the other hand
unique results and this problem is therefore removed. The inconsistencies
\cite{Ho:DySK87} mentioned above are also resolved in the sense that the
proposed reparametrization of time $x(t)=1-\ln t/\ln\bar t$ is not appropriate.

There has recently been a growing interest in the dynamics of glassy systems
evolving from a non-equilibrium initial state
\cite{CuKu93,FrMe94a,FrMe94b,CuKu94,BCuKuPa95,CuDe95,CuDou95}. The situation
there is more complicated because the problem is no longer homogeneous in time
and response and correlation functions depend on two time arguments $t$ and
$t'$. If, however,
$t-t'\ll t'$, one expects a behavior similar to the one investigated at
present with
$t'$ taking over the role of $\bar t$. Especially if both $t-t'$ and $t'$
diverge with $(t-t')/t'\to0$, one should also expect the appearance of
intermediate time scales and associated scaling regimes. This possibility has
not been considered at full depth in the above work.

For short time $t-t'$ correlation and response functions are related by
fluctuation-dissipation theorems  (FDT). This is also the case in the present
investigation. The different phases have, however, different scenarios of how
FDT's are violated at longer time scales. Since this happens for times $1\ll
t\ll \bar t$, similar phases are expected for the above mentioned relaxation
processes. In this context the non-equilibrium dynamics of the spherical
SK-model investigated recently by Cugliandolo and Dean \cite {CuDe95} is of
great interest, because this model does not require replica symmetry breaking
in its low temperature phase.

The present paper is organized as follows: Section 2 contains the definition
of the model and the dynamic mean field equations, which have the form of
coupled nonlinear integro-differential equations. It also contains the
definition of effective time dependent exponents, which are widely used in the
following. Section 3 is devoted to the numerical integration of the dynamic
mean field equations, to the presentation of results and to a preliminary
identification of different time scales and scaling regimes. Section 4
contains analytic results and estimates regarding time scales and scaling
properties. It starts with the FDT-solution valid for short time and continues
with a discussion of various convolution type integrals entering the
self-energies. In bypassing the QFDT-solution
\cite{KibachHo93I} and the hierarchical solution \cite{KibachHo93II} are
brought up. It is then shown how to evaluate the above integrals using the
time dependent effective exponents. This leads to a reformulation of the
original mean field equations in terms of a set of 15 coupled ordinary and
first order differential equations. This appears complicated but it allows a
discussion of different scaling regimes in which only closed subsets of these
equations are relevant. These regimes are in the order of increasing time the
FDT-regime, the plateau regime, where the correlation function
$q(t)$ stays close to the asymptotic value $q_c$ of the FDT-solution, and
finally the asymptotic regime for times of the order of the external time
scale $v^{-1}$. A concluding discussion follows in Section 5 and a brief
derivation of the dynamic mean field equations is given in an appendix.

\section{Formulation of the problem and dynamic mean field theory}
\setcounter{equation}{0}
%
\subsection{The model}

A particle is investigated which moves under the influence of an external force
$\bm{F}=\big\{\sqrt{N}F,0,\cdots,0\big\}$ in a random potential
$V(\bm{\varrho})$. The coordinates of the particle are written as
${\bm{\varrho}=\big\{\varrho_1+\sqrt{N}vt,\varrho_2,\cdots,\varrho_N\big\}}$
assuming a mean velocity $\bm{v}=\big\{\sqrt{N}v,0,\cdots,0\big\}$. This means
that the components $\varrho_i$ are measured in a frame moving with velocity
$\sqrt{N}v$ along the $1$-direction. The scaling with $\sqrt{N}$ ensures a
nontrivial limit
$N\to\infty$. The system is at a temperature $T=\beta^{-1}$ and its
Hamiltonian is
\begin{equation}
H=V(\bm{\varrho})+\sfrac12\mu_0\bm{\varrho}^2-\bm{F}\!\!\cdot\!\!\bm{\varrho}
\,,\end{equation} 
where a confining potential $\sfrac12\mu_0\bm{\varrho}^2$ might
be added to the random potential and the potential of the driving force. The
following investigation is, however, restricted primarily to $\mu_0=0$. The
motion of the particle is governed by a Langevin equation
\begin{equation}
\partial_t\varrho_{\alpha}=-\beta\frac{\delta H}{\delta
\varrho_{\alpha}}+\xi_{\alpha}
\end{equation} with Gaussian white noise
\begin{eqnarray}
\av{\xi_{\alpha}(t)\xi_{\beta}(t')}=2\,\delta_{\alpha\beta}\delta(t-t')\,.
\end{eqnarray} The quenched disorder is also assumed to be Gaussian with 
\begin{eqnarray} &&\overline{V(\bm{\varrho})}=0\nonumber\\
&&\overline{V(\bm{\varrho})V(\bm{\varrho'})}=-N\,f\big({\textstyle\frac 1 N}
(\bm{\varrho}-\bm{\varrho'})^2\,\big)
\label{VV-av}\end{eqnarray} and
\begin{equation} f(x)=\frac 1 {2(1-\gamma)}\,(1+x)^{1-\gamma}
\label{f-def}\,.\end{equation} The strength and the short range cutoff of the
disorder as well as the diffusion constant are set to one. This model differs
from the one studied earlier 
\cite{Mez:manifold90,Mez:manifold91,KibachHo93I,KibachHo93II} only by the
addition of the driving force.

\subsection{Dynamic mean field theory}

The order parameters of the dynamic mean field theory are the correlation
function
\begin{equation} q(t-t')=\frac 1
N\sum_{\alpha}\overline{\,\av{\big[\varrho_{\alpha}(t)
-\varrho_{\alpha}(t')\big]^2\,}\,}
\label{q-def}\end{equation} and the response function
\begin{equation} r(t-t')=\frac T N
\sum_{\alpha}\overline{\,\delta\av{\varrho_{\alpha}(t)} /\delta
F_{\alpha}(t')\Big.\,}\,.
\label{r-def}
\end{equation} They obey the mean field equations
\cite{KibachHo93I,KibachHo93II}
\begin{equation}
\partial_t r(t)=-\mu\,r(t)+\int_0^t \d{s}w(s)r(s)r(t-s)
\label{r-}\end{equation} and 
\begin{eqnarray}
\partial_t q(t)&=&2-\mu\,q(t)+\int_0^t \d{s}w(s)r(s)q(t-s)\nonumber\\
&-&\int_0^{\infty}\d{s}\Big\{2\big[W(t+s)-W(s)\big]r(s)\nonumber\\
&-&\big[w(t+s)r(t+s)-w(s)r(s)\big]q(s)\Big\}\hspace{1.4cm}
\label{q-}\end{eqnarray} with
\begin{equation} w(t)=-4\beta^2f''\big(\,q(t)+v^2t^2\,\big)\,,
\label{w-def}\end{equation}

\begin{equation} W(t)=2\beta^2f'\big(\,q(t)+v^2t^2\,\big)
\label{W-def}\end{equation} and 
\begin{equation}
\mu=\mu_0+\int_0^{\infty}\d{s}w(s)r(s)\,.
\label{mu-def}\end{equation} A brief derivation is given in the appendix. In
the following $\mu_0=0$ is assumed. 

In thermal equilibrium correlation and response functions are related by a
fluctuation-dissipation theorem (FDT) which reads $\partial_t q(t)=2r(t)$.
This is violated in a non-equilibrium situation and
\begin{equation} n(t)=\frac{1}{2r(t)}\,\partial_tq(t)-1
\label{n-def}\end{equation} is introduced as measure of this FDT-violation.

Using this in \citeq{w-def} and \citeq{W-def}
\begin{equation}
\partial_t W(t)=-\Big\{\big[1+n(t)\big]r(t)+v^2t\Big\}w(t)
\label{dW-}\end{equation} is found. Combining \citeq{r-}, \citeq{q-} and
\citeq{n-def} one obtains
\begin{eqnarray} r(t)\partial_tn(t)&=&-\int_0^t \d{s}w(s)r(s)r(t-s)\nonumber\\
&&\hskip20pt\times\big[n(t)-n(t-s)\big]\nonumber\\
&&+\int_0^{\infty}\d{s}\Big\{w(t+s)r(t+s)r(s)\nonumber\\
&&\hskip20pt\times\big[n(t+s)-n(s)\big]\nonumber\\
&&\hskip20pt+\,v^2\big(t+s\big)\,w(t+s)r(s)\Big\}\hspace{1.4cm}
\label{n-}\end{eqnarray} and this equation can now be used instead of one of
the original mean field equations
\citeq{r-} or \citeq{q-}.

Prescribing the drift velocity $v$ the average of the force necessary to drive
the particle is
\begin{equation}
\beta \,F=v\,\Big\{1+\int_0^{\infty}\d{s}s\,w(s)r(s)\Big\}
\label{F-}\end{equation} which is derived in the appendix.

\subsection{Effective exponents}

It is convenient to define 
\begin{equation} g(t)=t\,r(t)
\label{g-}\end{equation} and then \citeq{n-def} is written as
\begin{equation} t\partial_t q(t)=2\big\{1+n(t)\big\}g(t)\,.
\label{dq-}\end{equation}

In order to present the results of the numerical integration of the mean field
equations effective time dependent exponents are introduced. They are also
used in the evaluation of the convolution type integrals in the self-energies
of the mean field equations \citeq{r-} and \citeq{n-}. The first two exponents
are
\begin{equation}
\nu(t)=t\partial_t\ln g(t)
\label{nu-def}\end{equation} and
\begin{equation}
\alpha(t)=-t\partial_t \ln w(t)\,.
\label{alpha-def}\end{equation} With
\begin{equation} k(t)=t\partial_t n(t)
\label{k-def}\end{equation} the third exponent
\begin{equation}
\kappa(t)=t\partial_t\ln k(t)
\label{kappa-def}\end{equation} is defined.

\section{Numerical results}
\setcounter{equation}{0}
%
\subsection{Numerical procedure}

The dynamic mean field equations to be solved consist of two
integro-differential equations \citeq{r-}and \citeq{n-}, a first order
differential equation
\citeq{n-def}, the definition of $w(t)$, \citeq{w-def} and \citeq{f-def}, and
the integral \citeq{mu-def} determining $\mu$. The second integral in
\citeq{n-} and the integral \citeq{mu-def} extend overall times and therefore,
evaluating the solution at some $t$ requires the knowledge of the solution at
all times, not only at $s<t$. This means that the equations have to be
iterated. This is the price one has to pay for investigating a steady state
situation. A study of the relaxation from a non-equilibrium initial state on
the other hand is free of this problem, but it requires to deal with functions
depending on two time arguments.

The wide span of time arguments ranging from $10^{-4}$ to $10^{36}$ requires a
nonuniform discretization. The present calculation uses a homogeneous
discretization of $\ln t$. The convolution type integrals in \citeq{r-} and
\citeq{n-} can be evaluated by numerical integration using the values stored
at the points of the above grid for one of the functions of the integrand and
an interpolation for the other function. Alternatively, an approximative
evaluation of the integrals using effective exponents, described in Section
4.4 is possible. For the actual calculation both methods have been combined.

\mfigure{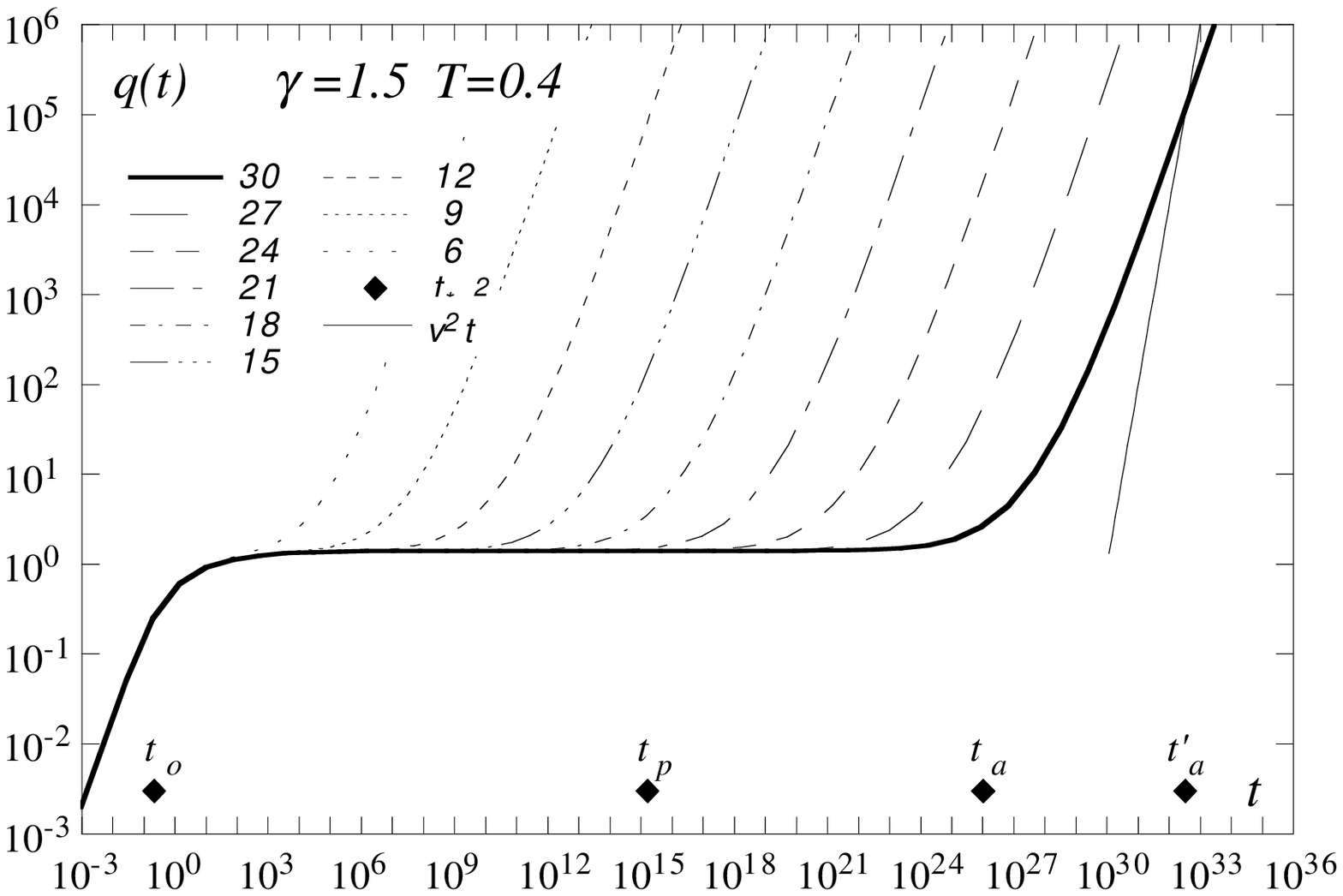} {Correlation function $q(t)$ in phase $\cal A$}{q-sr}

\mfigure{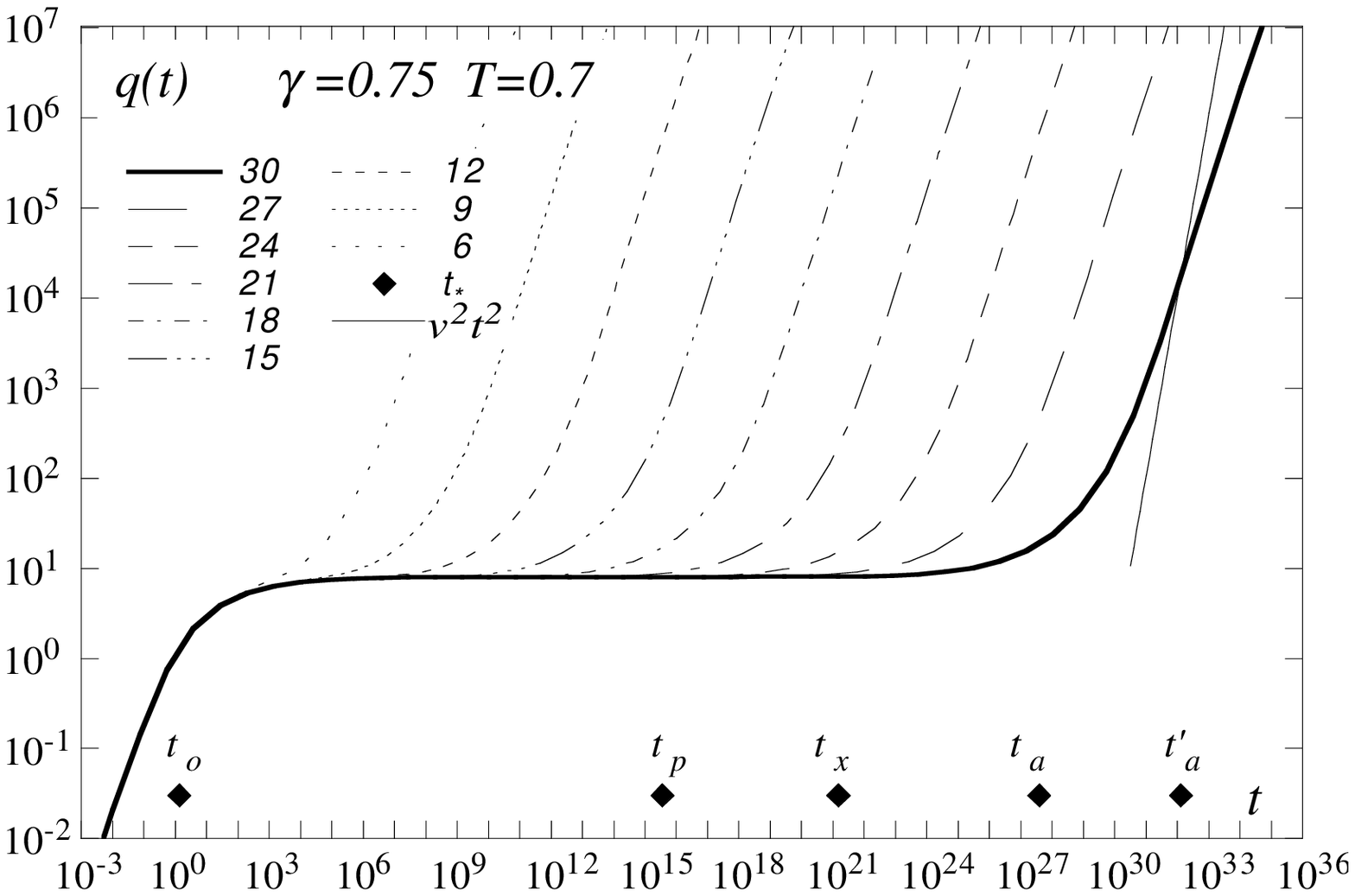} {Correlation function $q(t)$ in phase $\cal B$}{q-lr}

The mean field equations in the form written in Section 4 consist of several
coupled nonlinear differential equations of first order. They can be
integrated forward or backward in time. In addition they have to be iterated.
The situation resembles in some sense the problem of solving 
\begin{equation}
\frac {\;\d x(t)}{\;\d t}=f\big(x(t)\big)\,.
\end{equation} Integrating forward in $t$ the solution typically approaches a
fixed point 
$f(\bar x)=0$ with $f'(\bar x)<0$, whereas a fixed point with $f'(\bar x)>0$ is
reached integrating backward in $t$. Selecting an appropriate initial value
$x(t_0)$ somewhere between a fixed point with $f'(\bar x)>0$ and an adjacent
one with 
$f'(\bar x)<0$, the solution can easily be found by integration in both
directions. For the complete set of equations such fixed point situations show
up at the crossover between the various scaling regimes discussed below.

\subsection{Results}



\mfigure{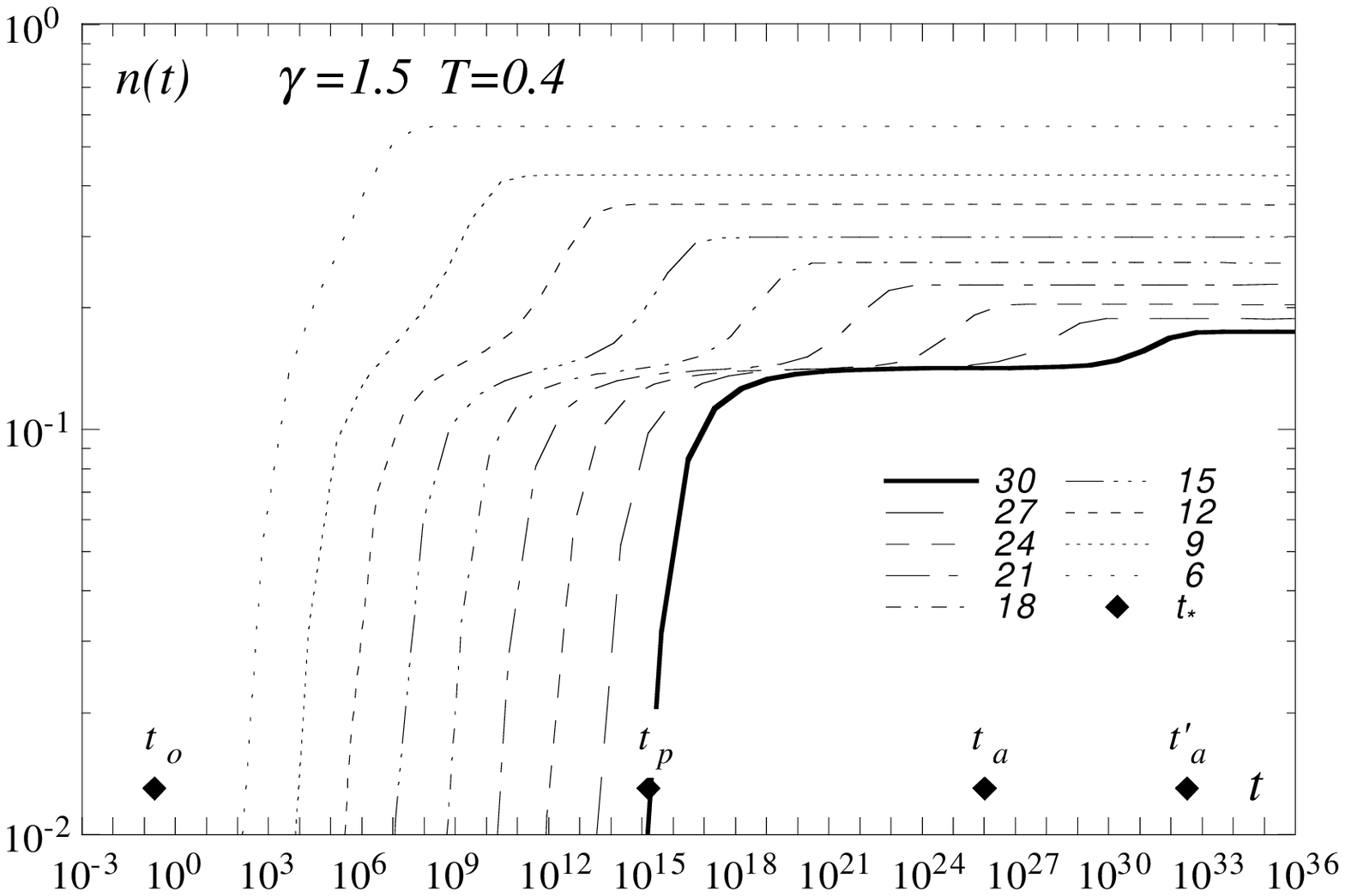} {$n(t)$ in phase $\cal A$}{n-sr}

\mfigure{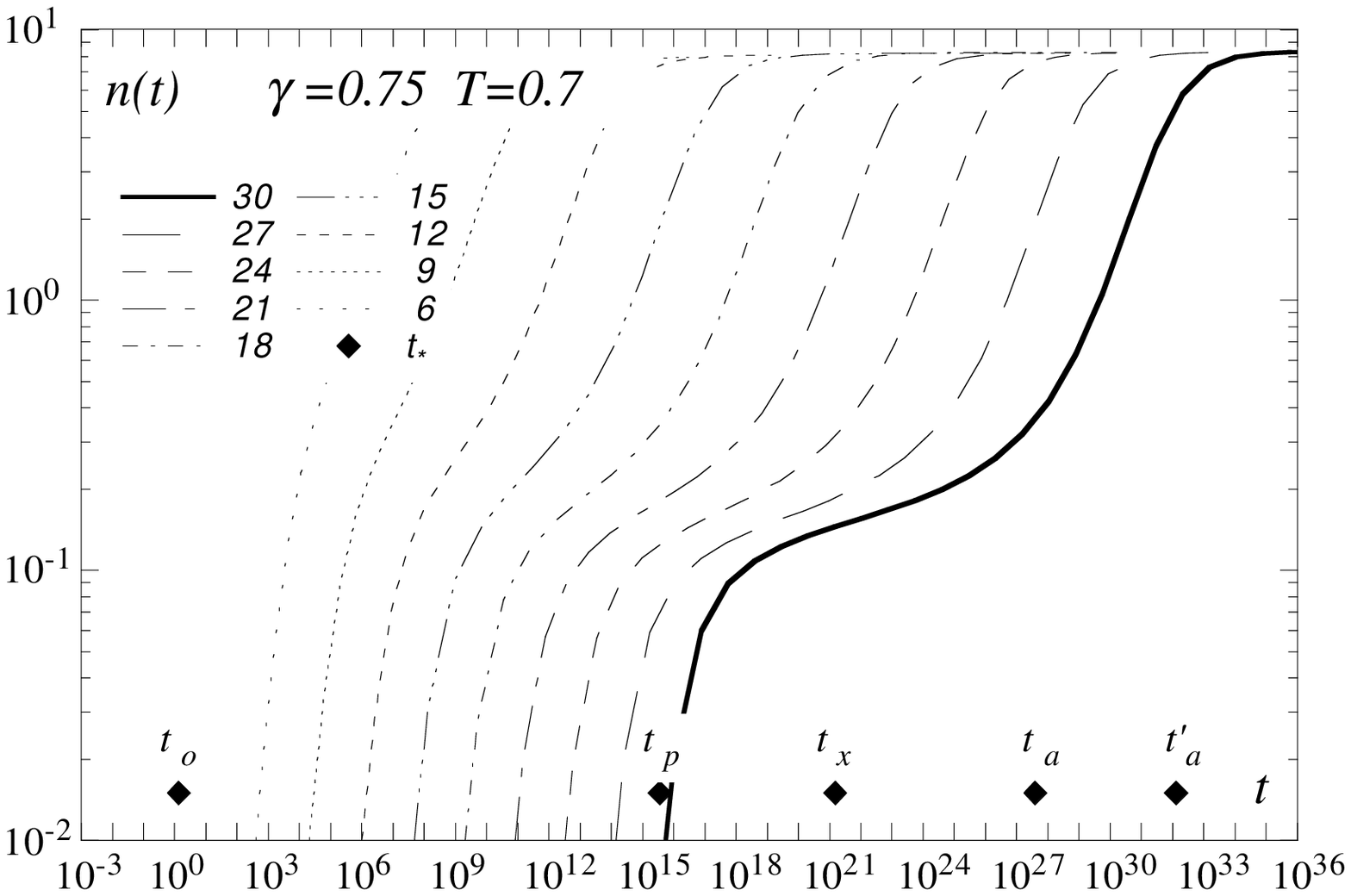} {$n(t)$ in phase $\cal B$}{n-lr}

\mfigure{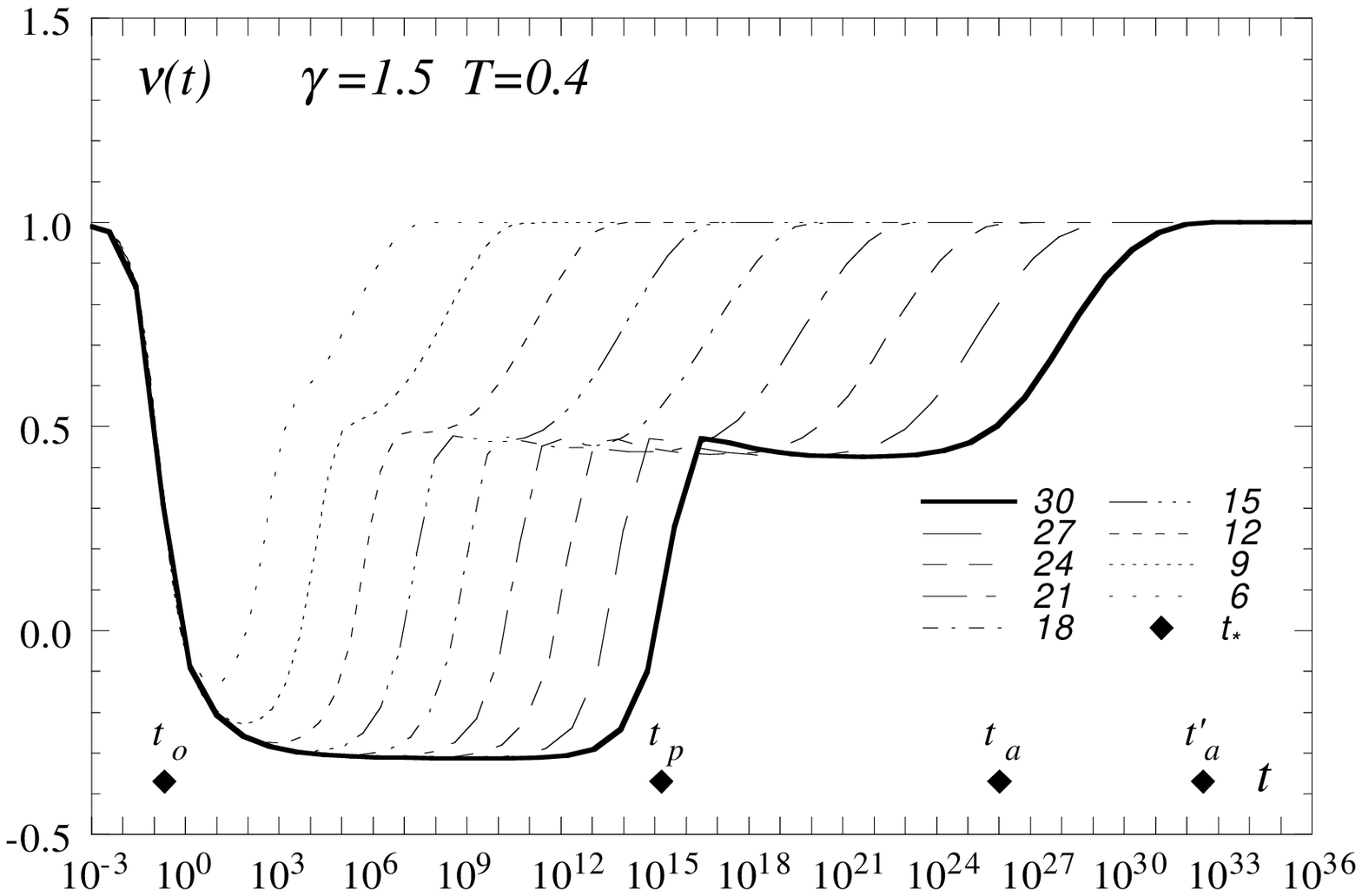} {Effective exponent $\nu(t)$ in phase $\cal A$}{nu-sr}

\mfigure{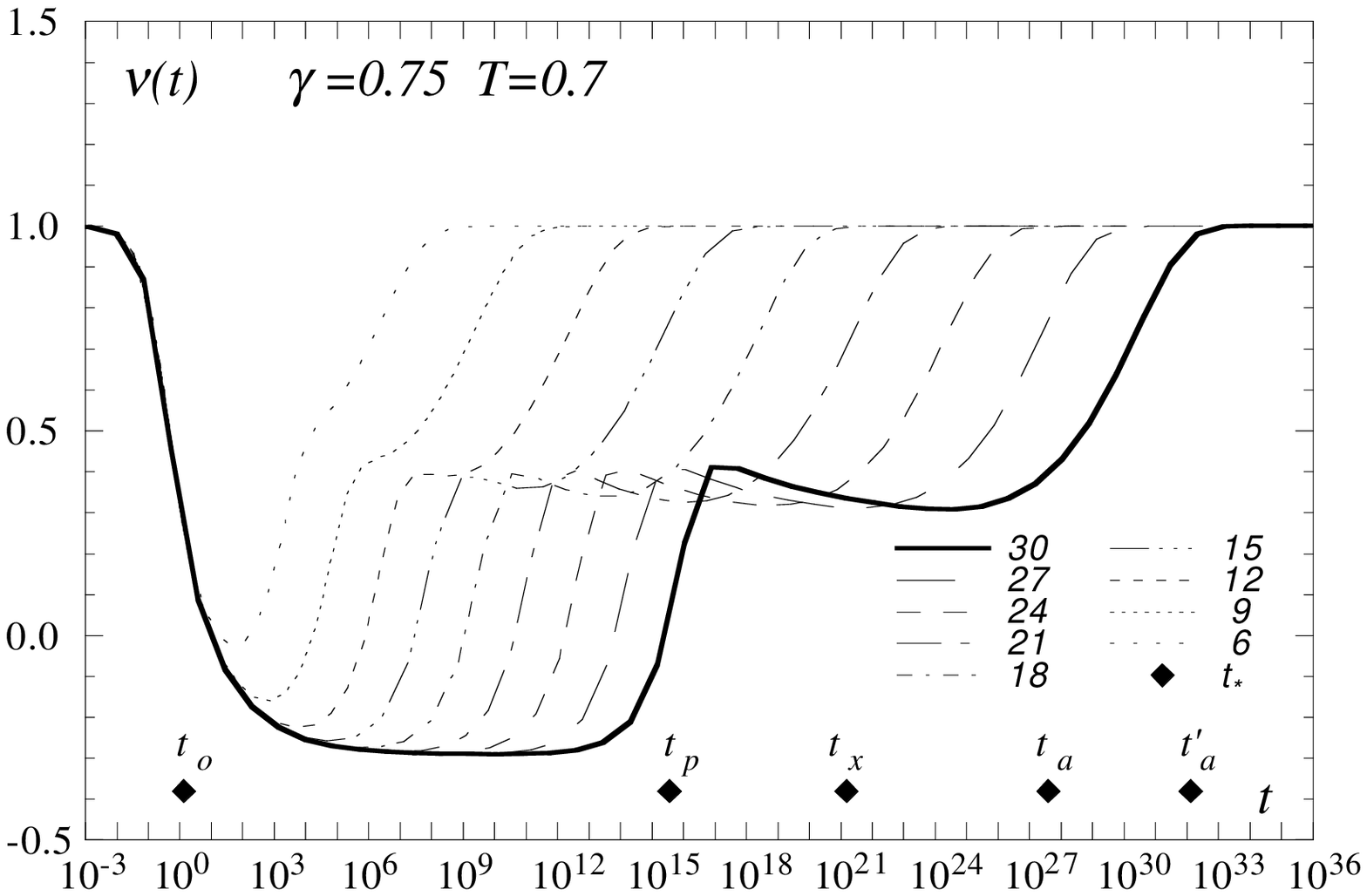} {Effective exponent $\nu(t)$ in phase $\cal B$}{nu-lr}

\mfigure{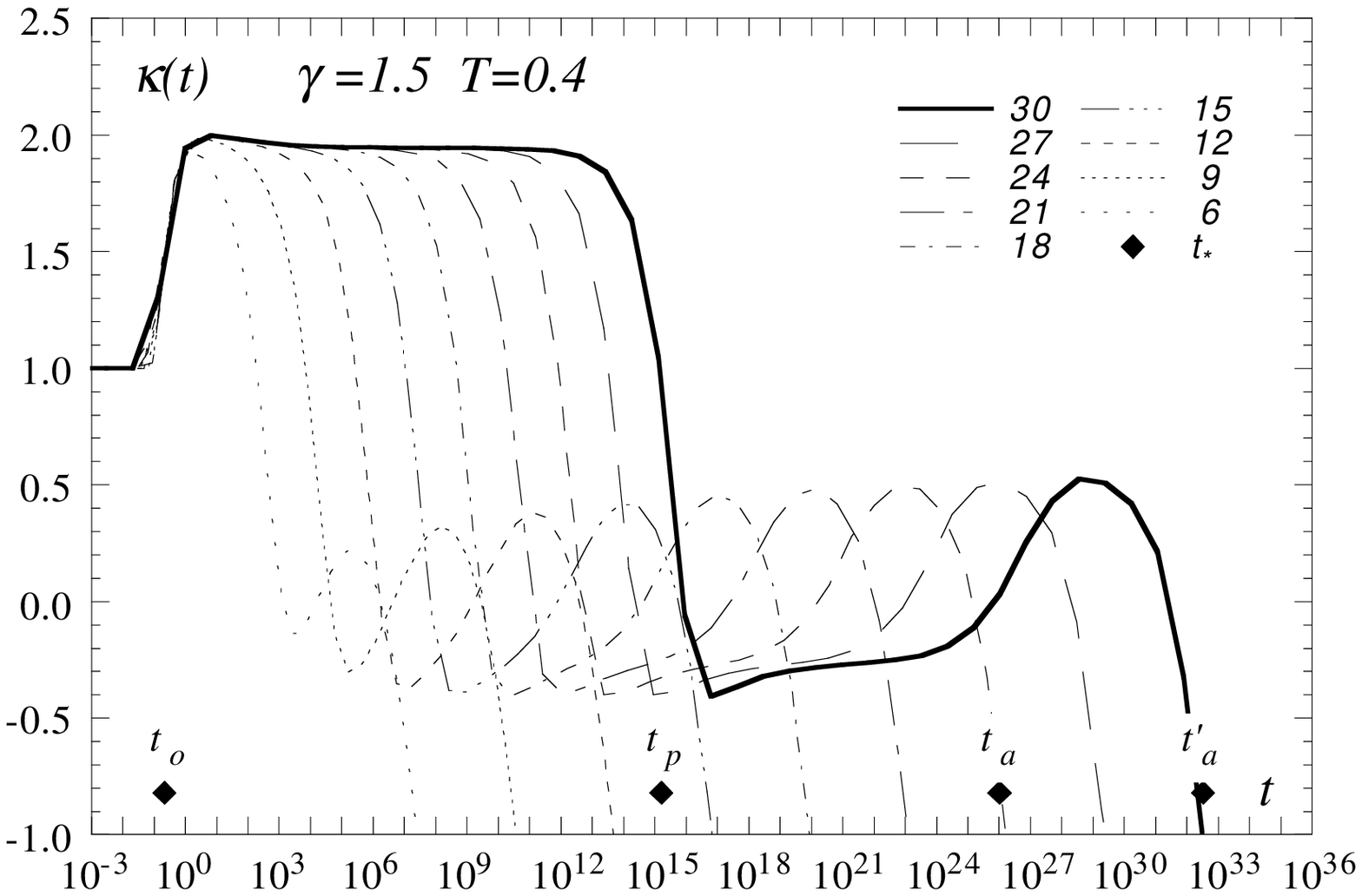} {Effective exponent $\kappa(t)$ in phase $\cal A$}{ka-sr}

\mfigure{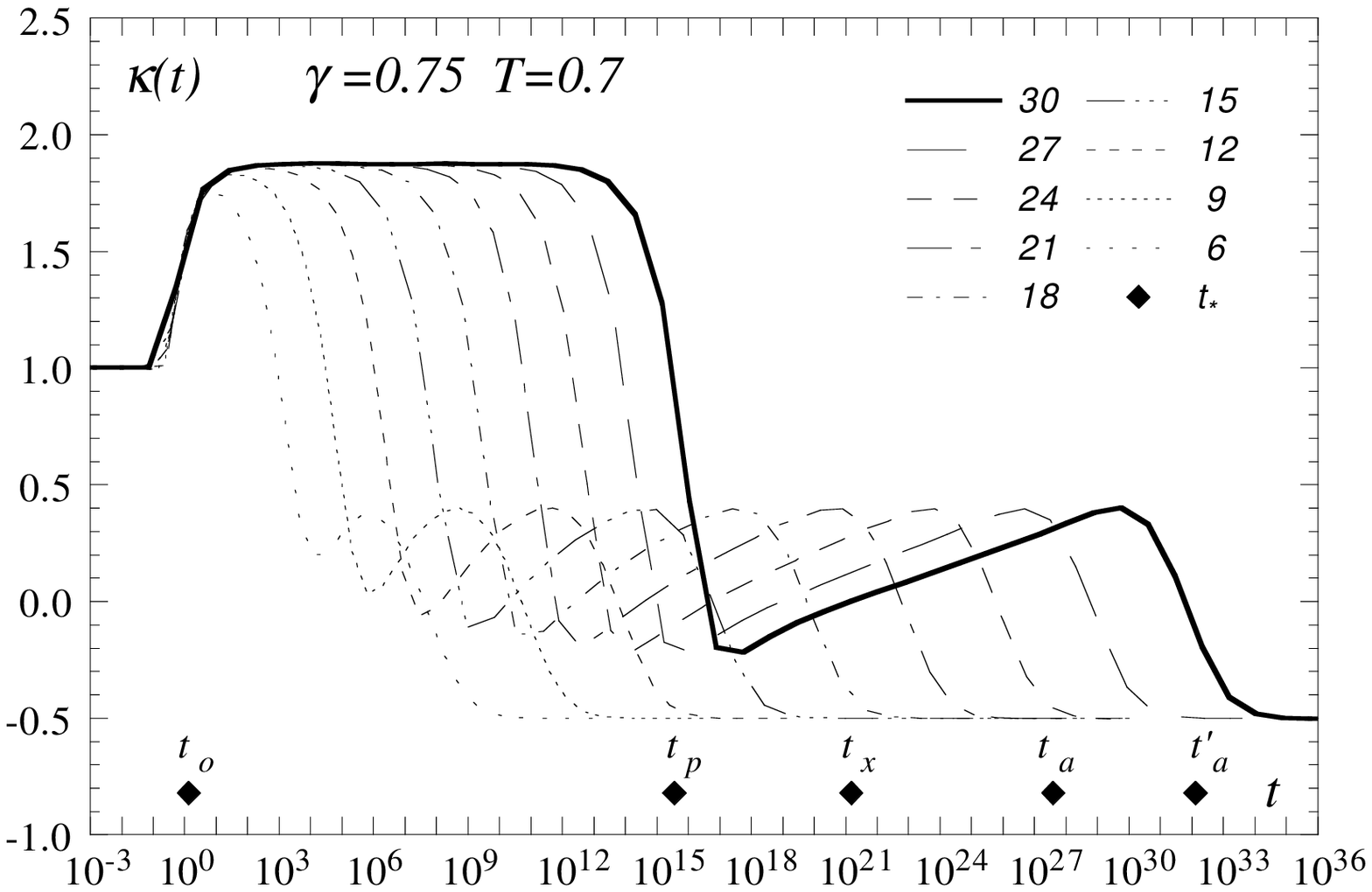} {Effective exponent $\kappa(t)$ in phase $\cal B$}{ka-lr}

\mfigure{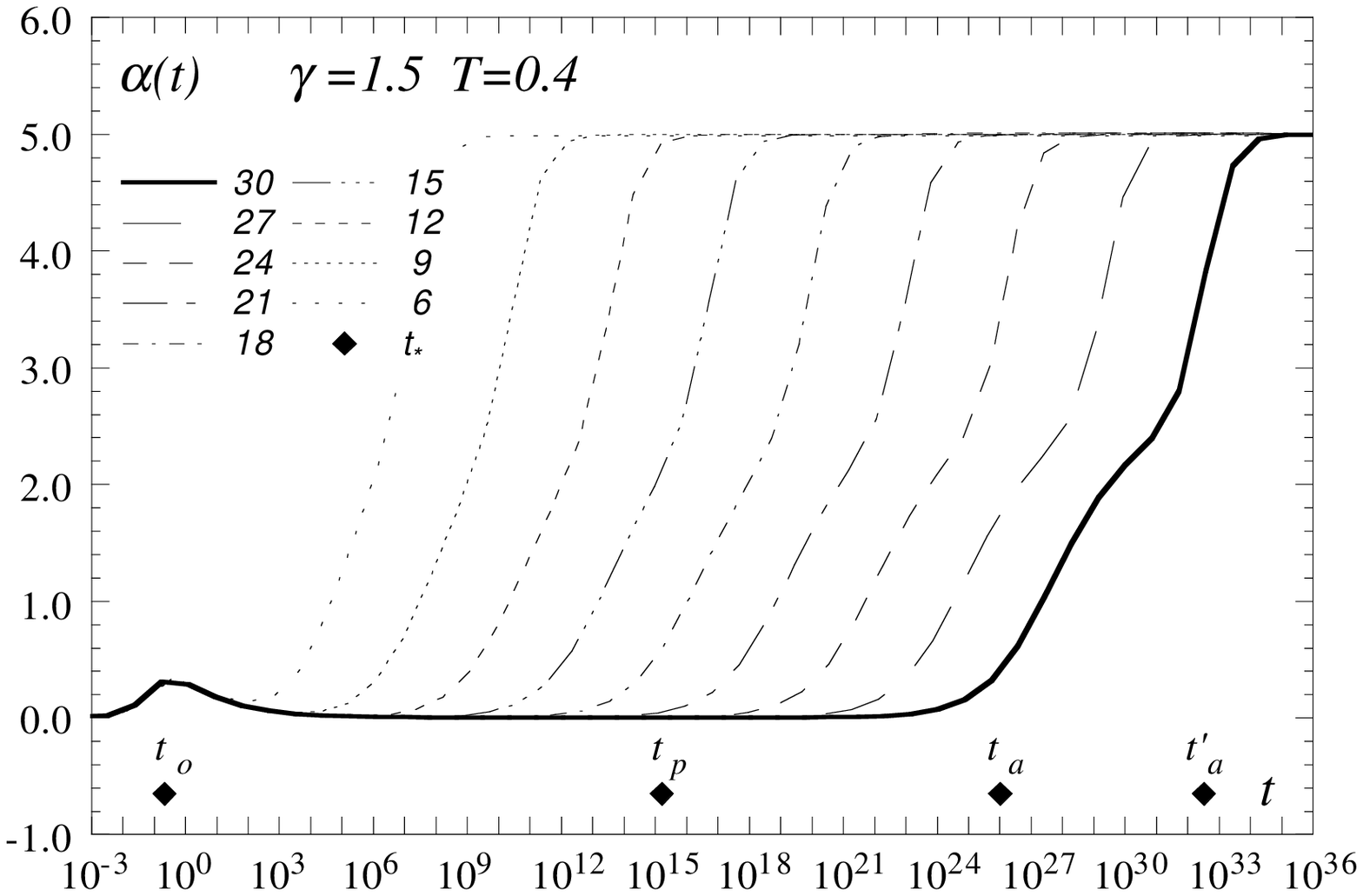} {Effective exponent $\alpha(t)$ in phase $\cal A$}{al-sr}

\mfigure{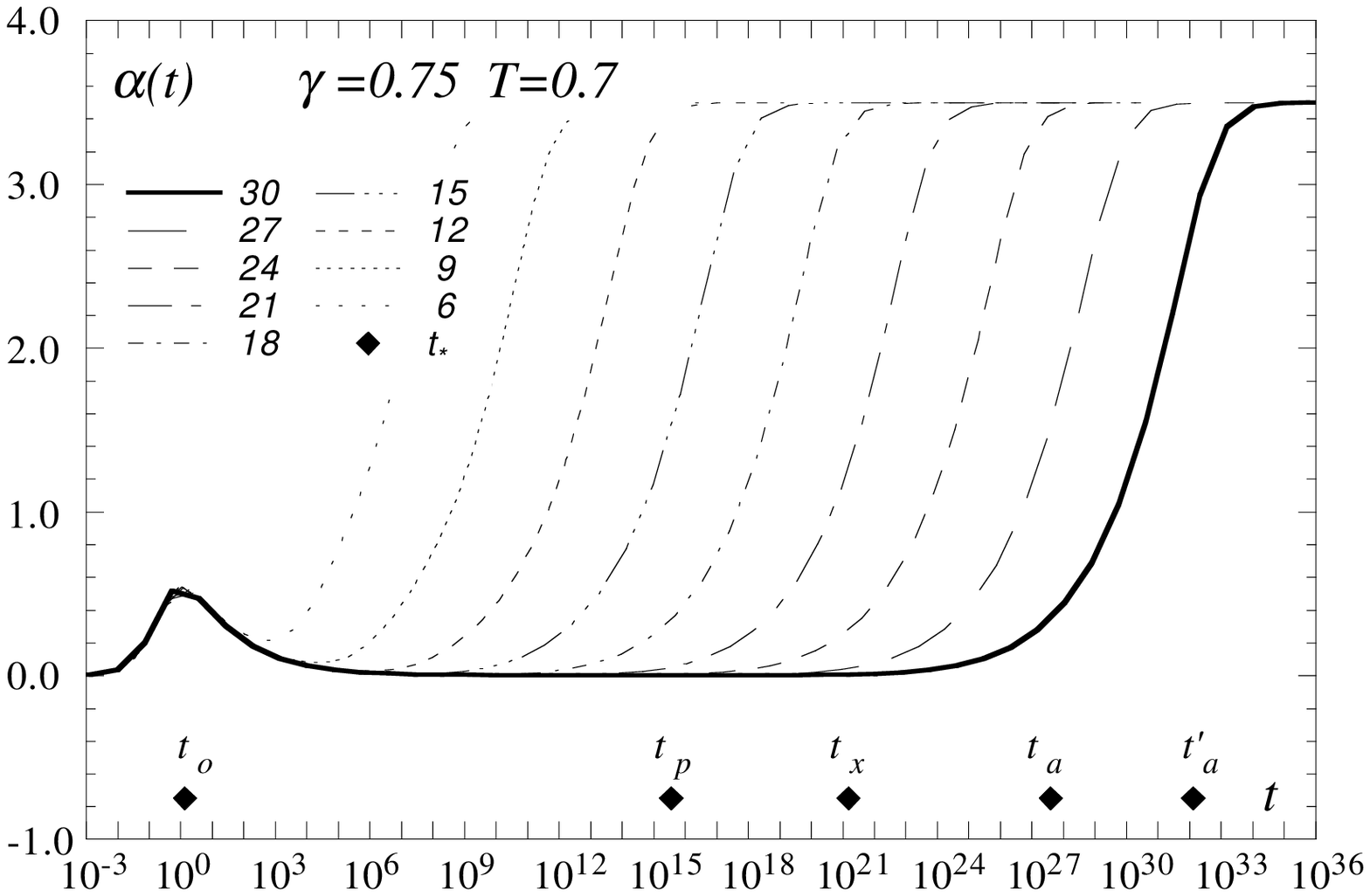} {Effective exponent $\alpha(t)$ in phase $\cal B$}{al-lr}

\mfigure{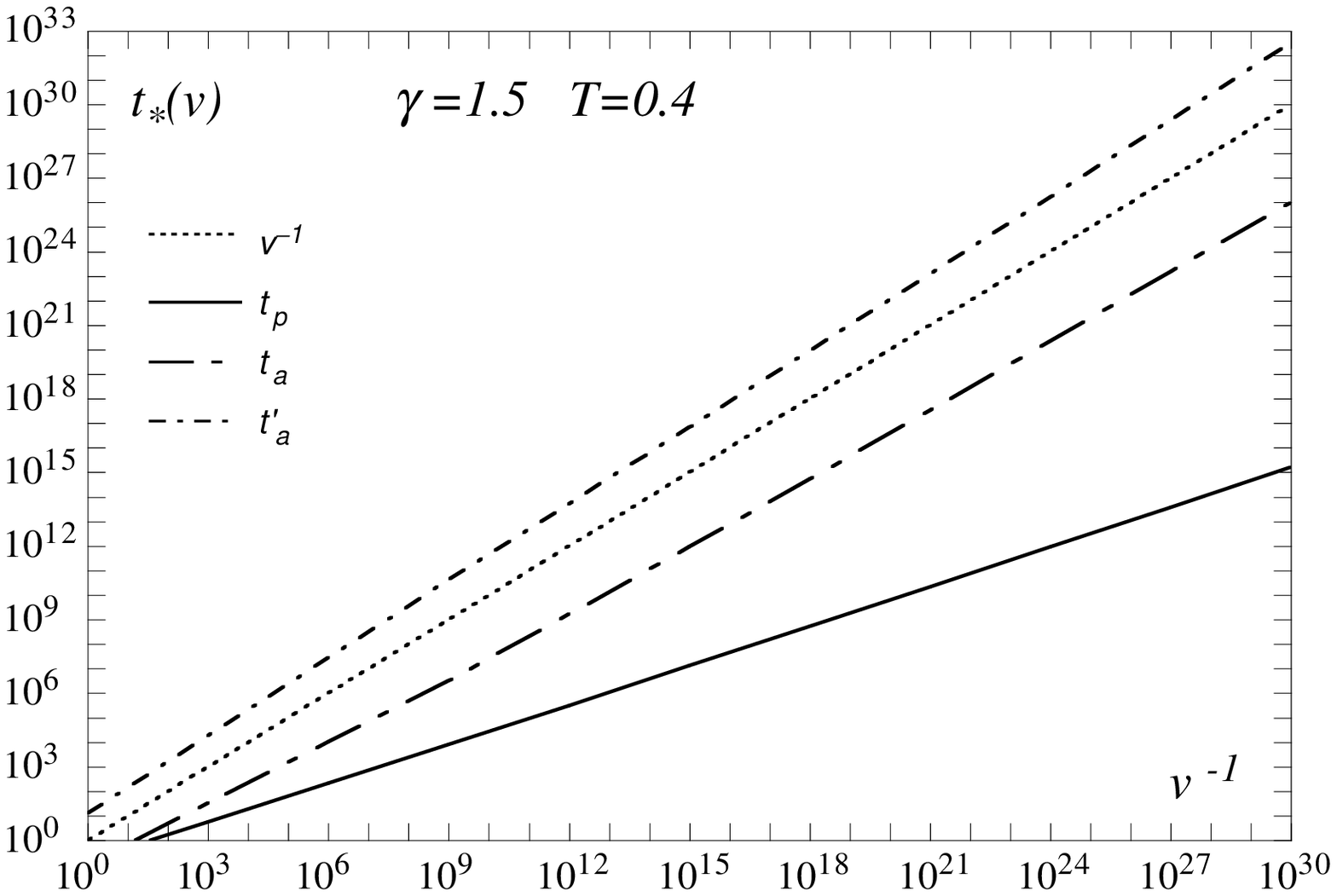} {Characteristic time scales  in phase $\cal A$}{tv-sr}

\mfigure{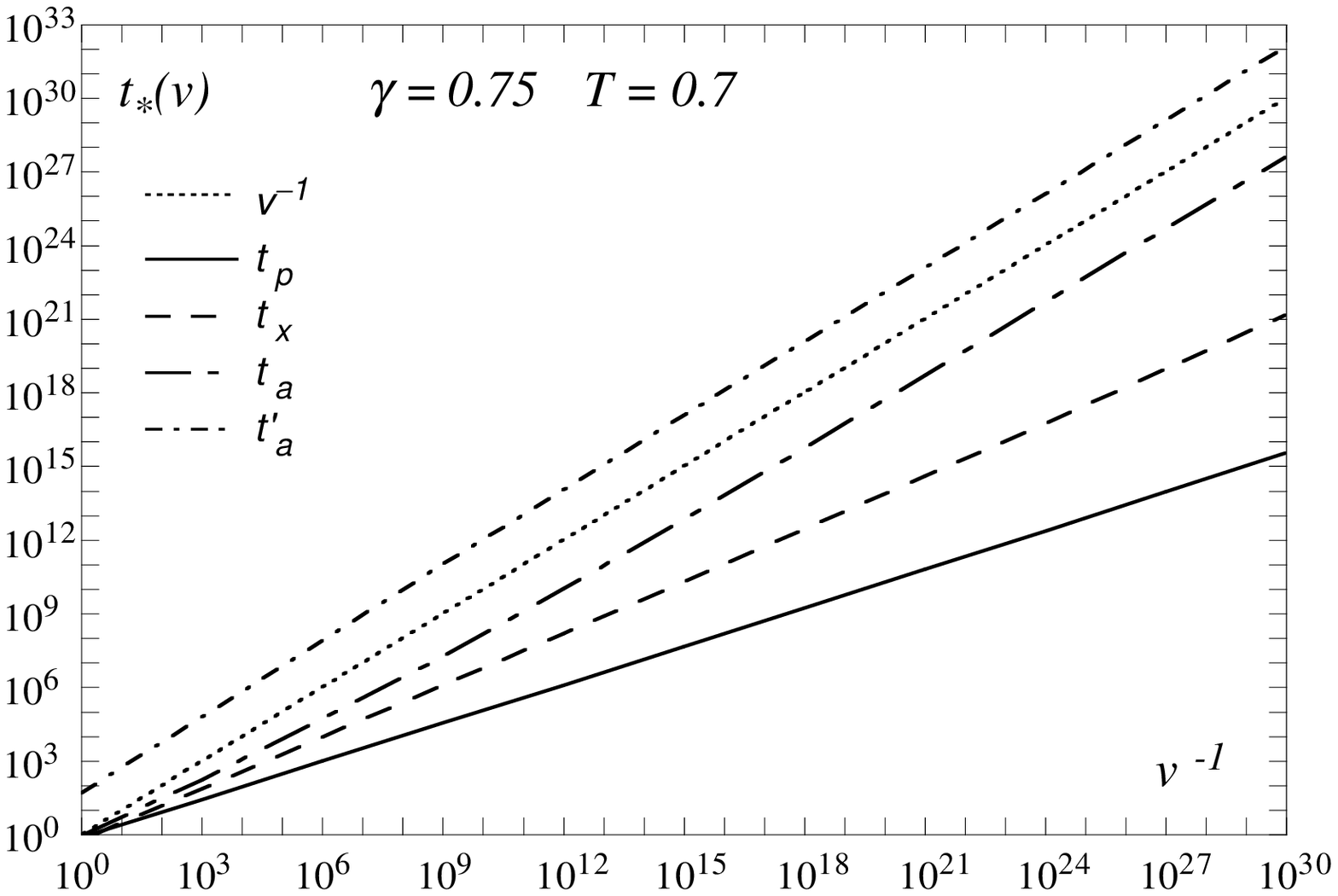} {Characteristic time scales  in phase $\cal B$}{tv-lr}

The following figures show selected results obtained for $\{\gamma=1.5;\;
T=0.4\}$ and
$\{\gamma=0.75;\; T=0.7\}$, respectively. The first set of variables
corresponds to a situation where the replica calculation
\cite{Mez:manifold90,Mez:manifold91} requires single step replica symmetry
breaking. This corresponds to a point in phase $\cal A$ of the phase diagram,
Fig.\ref{phase-fig}. The second set corresponds to a point in phase $\cal B$
and the corresponding replica treatment requires continuous replica symmetry
breaking. The following figures show results obtained for various values of
the drift velocity ranging from $v=10^{-6}$ to $v=10^{-30}$. In the following
various velocity dependent characteristic time scales are introduced. Their
values for $v=10^{-30}$ are marked in the figures.

Figs.\ref{q-sr} and \ref{q-lr} show the correlation function $q(t)$. The
plateau value $q_c$ is given by \citeq{T-FDT}. In view of the argument
$q(t)+v^2t^2$ in
\citeq{w-def} the value of $v^2t^2$ for $v=10^{-30}$ is also shown.

The function $n(t)$ defined in \citeq{n-def} is shown in Figs.\ref{n-sr} and
\ref{n-lr}. It indicates the violation of the FDT, which holds for
$t<t_p$. It develops a first plateau value for $t_p<t<t_a$ and a second one for
$t>t'_{a}$. In phase $\cal A$ this second value obviously depends on $v$. In
phase
$\cal B$ the first plateau is not yet fully developed, even for $v=10^{-30}$.

The time dependent effective exponents $\nu(t)$, \citeq{nu-def}, $\kappa(t)$,
\citeq{kappa-def}, and $\alpha(t)$, \citeq{alpha-def}, are shown in
Figs.\ref{nu-sr} to \ref{al-lr}. Their properties are discussed below.

From the data several characteristic time scales can be extracted. They are
shown in Figs.\ref{tv-sr} and \ref{tv-lr} as functions of the velocity $v$.
The time $t_p$ is defined by $q(t_p)=q_c$ and it characterizes the center of
the plateau. The next time scale $t_x$ is relevant only for phase $\cal B$ and
it is given by
$\kappa(t_x)=0$.  The time scale $t_a$ is defined by $q(t_a)=2\,q_c$ and
$t_a'$ by
$q(t_a')=2\,v^2{t_a'}2$. A power law dependence is found for $v\to0$.

\subsection{Scaling regimes}

The numerical results and analytic considerations described in Section 4
reveal the existence of various scaling regimes in the limit $v\to0$. 

\subsubsection{FDT-regime} For finite $t\sim t_0$ one finds $n(t)\ll1$ and
fluctuation dissipation theorems hold. This can be understood as a situation
where the particle stays within one valley of the energy landscape. The
correlation function and $n(t)$ can be written as
\begin{eqnarray} q(t)&=&\otop q(t/t_0)\nonumber\\ n(t)&=&\otop b(v)\,\otop
n(t/t_0)
\label{scf-FDT}\end{eqnarray} with $\otop b(v)\to0$ for $v\to0$. The time
scale $t_0$ and the functions $\otop q(x)$ and $\otop n(x)$ do not depend on
$v$. This regime is referred to as the FDT-regime.

\subsubsection{Plateau-regime} With increasing $t$ a plateau regime is found
where $q(t)\approx q_c$. The corresponding time scale is $t_p=t_p(v)$ defined
by $q(t_p)=q_c$ and the above functions obey the scaling form
\begin{eqnarray} q(t)&=&q_c+\hat a(v)\,\hat q(t/t_p)\nonumber\\
n(t)&=&n_c+\hat n(t/t_p)
\label{scf-pl}\end{eqnarray} which is consistent with $\nu(t)=\hat\nu(t/t_p)$
and
$\kappa(t)=\hat\kappa(t/t_p)$.

The function $n(t)$, which is a measure of FDT-viola\-tion, reaches a plateau
value
$n_c$ for $t\sim t_p$. In phase $\cal A$ this plateau extends beyond the
limits of the
$q$-plateau. 

In phase $\cal B$ the plateau of $n(t)$ is not yet very pronounced, even at
$v=10^{-30}$, but it can be seen that the upper limits of the
$n$-plateau and the $q$-plateau essentially coincide. The center of the
$n$-plateau defines an additional time scale $t_x=t_x(v)$ where $n(t_x)=n_c$
or $\kappa(t_x)=0$. For $t\sim t_x$ a scaling form
\begin{equation} n(t)=n_c+c(v)\hat n_x\big(c(v)\ln t/t_x)\big)
\label{n-x}\end{equation} will be derived.

In phase $\cal C$ $\;n(t)\ll1$ holds up to the upper end of the $q$-plateau and
consequently the FDT-solution holds in the whole plateau regime.

\subsubsection{Asymptotic regime}

The lower end of the asymptotic regime is marked by a time scale $t_a=t_a(v)$
where
$q(t_a)-q_c\sim q_c$. An appropriate choice is $t_a$ such that $q(t_a)=2q_c$.
For
$t\sim t_a$ a scaling form
\begin{eqnarray} q(t)&=&\bar q(t/t_a)\nonumber\\ n(t)&=&n_c+\bar b(v)\bar
n(t/t_a)
\label{scf-as}\end{eqnarray} with $\bar b(v)=1$ in phase $\cal B$ and $\cal C$
is found.

Another time scale $t'_a= t'_a(v)$ can be defined such that
$q(t'_a)=\big(v\,t'_a\big)^2$. In phase $\cal B$ and $\cal C$ both are
proportional to
$v^{-1}$, whereas $t_a(v)\sim v^{-1+\eta}$ and $t'_a(v)\sim v^{-1-\eta}$ with
$\eta>0$ is found in phase $\cal A$. This requires for $t\sim t'_a$ the
scaling form
\begin{equation} n(t)=n_c+\bar b'(v)\,\bar n'(t/t_a ')\,.
\label{scf-as'}\end{equation} The existence of the exponent $\eta>0$ is
connected to the creep behavior observed in phase $\cal A$.

The above results now have to be verified by analytic investigations. This is
done in the following sections.

\section{Analytic results}
\setcounter{equation}{0}
%
\subsection{FDT-solution}

The shortest $v$-dependent time scale for $v\to0$ is $t_p$ with
${1\ll t_p\ll v^{-1}}$. For $t\ll t_p$ one expects a solution which obeys the
FDT and therefore $n(t)=0$. This solution holds in phase $\cal O$ for $v=0$
and describes equilibrium within a single valley of the energy landscape in
the other phases. The discussion of this solution follows standard arguments
\cite{KibachHo93I}. With
\citeq{q-},
\citeq{dW-} and \citeq{dq-} one obtains
\begin{eqnarray}
\partial_tq(t)&=&2-\mu q(t)\nonumber\\&&+\int_0^t\d{s}w(s)r(s)q(t-s)
\label{dq-FDT}\,,\end{eqnarray} because the second integral in \citeq{q-}
vanishes. In order to demonstrate this, one realizes that its main
contributions come from $s\sim t$ where $n(t)\approx 0$. With \citeq{g-} and
\citeq{dq-} $\partial_t q(t)\approx 2r(t)$  and with \citeq{dW-}
$\partial_t W(t)\approx-w(t)r(t)$. This allows to write the integrand as a
complete derivative with respect to $s$ and to evaluate the integral resulting
in $0$, since $q(0)=0$ and $W(t)\to 0$  for $t\to\infty$.

Eqs.\citeq{r-} and \citeq{dq-FDT} cannot be solved in closed form. It is,
however, sufficient to investigate the solution for $t\to t_p$, where $q(t)\to
q_c$, $\partial_t q(t)\to0$ and $\partial_t r(t)\to0$.

In leading order the integrals in both equations can be evaluated by taking
into account only the contributions near the upper and lower bound,
respectively. This yields from \citeq{r-}
\begin{equation}
\mu=2\beta^2\big\{f'(0)-f'(q_c)-q_cf''(q_c)\big\}
\label{mu-FDT}\end{equation} and from \citeq{dq-FDT}
\begin{equation} f''(q_c)q_c^2=-T^2\,.
\label{T-FDT}\end{equation} The second equation is used to determine $q_c$.
Elimination of $\mu$ in
\citeq{mu-FDT} using \citeq{mu-def} yields with \citeq{w-def},\citeq{W-def} 
and \citeq{dq-}
\begin{eqnarray}
q_cf''(q_c)&=&-\frac{1}{2\beta^2}\int_{t_p}^{\infty}\d{s}w(s)\,r(s)\nonumber\\
&=&\int_{t_p}^{\infty}\d{s}\dot q(s)\frac{f''\big(q(s)+v^2s^2\big)}{1+n(s)}\,.
\label{condition}\end{eqnarray} This condition involving the solutions at long
time scales will be used later.

In order to give estimates of the corrections to the leading order one assumes
\begin{equation} t\,r(t)=g(t)\toto_{t\to t_p}
-\sfrac12\nu_0q_c\,(t/t_0)^{\nu_0}
\label{g-FDT}\end{equation} with $\nu_0<0$. The time scale $t_0$ has yet to be
determined. This leads to
\begin{eqnarray} q(t)&&=q_c-2\int_t^{t_p}\d{s}r(s)\nonumber\\
&&\toto_{t\to t_p}q_c\Big\{1-(t/t_0)^{\nu_0}\Big\}\,.
\label{q-FDT}\end{eqnarray}

This is now used in an analysis of the corrections in \citeq{r-} where the
leading order contains terms $\sim t^{\nu_0-1}$. The next to leading order 
$\sim t^{2\nu_0-1}$ results in
\begin{eqnarray}
&&f''(q_c)\int_0^t\d{s}\big\{r(s)r(t-s)-2r(t)r(s)\big\}\nonumber\\
&&\qquad\quad =\big\{f''(q_c)+\sfrac12 f'''(q_c)q_cr(t)\big\}
\big\{q_c-q(t)\big\}\,.\qquad
\end{eqnarray}

With \citeq{g-FDT} the integral on the left hand side can be evaluated.
Introducing
\begin{equation} R_m=-\frac{q_c\,f'''(q_c)}{2f''(q_c)}
\label{R-def}\end{equation} one obtains
\begin{equation}
\frac{\Gamma^2(1+\nu_0)}{\Gamma(1+2\nu_0)}=R_m
\label{nuo-}\end{equation} which allows to determine the exponent $\nu_0$.

In order to get an estimate of the time scale $t_0$ one can use an
interpolation formula
\begin{equation} q(t)\approx q_c\Big\{1-\big(1+t/t_0\big)^{\nu_0}\Big\}
\end{equation} which gives the correct asymptotic behavior for $t\to t_p$. The
requirement $q(t)\to 2t$ for $t\to0$ yields 
\begin{equation} t_0\approx-\sfrac12\nu_0q_c\,.
\label{to-}\end{equation}

In the ergodic high temperature phase $\cal O$ for $v=0$ the FDT holds for all
times and therefore with \citeq{mu-def} 
\begin{equation}
\mu=\frac{2f'(0)}{T^2}\,.
\end{equation} At the transition to one of the nonergodic phases the plateau
develops and one finds with \citeq{mu-FDT} and \citeq{T-FDT}
\begin{eqnarray} T_c^2&=& q_cf'(q_c)\,,\nonumber\\ f'(q_c)&=&-q_cf''(q_c)\,.
\label{Tc-}\end{eqnarray} A solution of these equations with finite $T_c$
exists only for $\gamma>1$. It is 
\begin{eqnarray} q_c&=&\frac1{\gamma-1}\nonumber\\
T_c&=&\sqrt{\sfrac12}\,\gamma^{-\gamma/2}\,(\gamma-1)^{(\gamma-1)/2}\,.
\label{Tc--}\end{eqnarray} This is shown in Fig.\ref{phase-fig}. The
corresponding value obtained from replica theory
\cite{Mez:manifold90,Mez:manifold91}
\begin{equation} T_{c,\rm1RSB}=\frac{1}{\sqrt{6\gamma}}
\label{Tc-1RSB}\end{equation} is always below the above value.

For $T>T_c$ in phase $\cal O$ the FDT holds for all times and the force
\citeq{F-} is 
\begin{equation}
\beta\,F=v\,\Big\{1+\int_0^{\infty}\d{t}W(t)\Big\}\,.
\label{Foo}\end{equation} For $t\to \infty$ \citeq{r-} yields $r(t)\to
r_{\infty}$ and $q(t)\sim t$. Therefore the integral in \citeq{Foo} converges
for $\gamma>1$ and there exists a finite friction constant for $v\to 0$

\subsection{Integrals and counterterms}

For $t\gg1$ the leading contributions to the integrals in \citeq{r-} and
\citeq{n-} come from regions near the boundaries of integration. Subtracting
those leads to residual integrals and a reformulation of the dynamic mean
field equations.  The following integrals are introduced:
\begin{eqnarray}
J(t)&=&\int_0^t\d{s}\Big\{w(s)r(s)r(t-s)-w(t)r(t)r(t-s)\nonumber\\
&&-w(s)r(s)r(t)+s\,w(s)r(s)\partial_t r(t)\Big\}\,,\quad\quad
\label{J-def}\end{eqnarray}
\begin{eqnarray} K(t)&=&\int_0^{\infty}\d{s}w(t+s)r(t+s)r(s)\nonumber\\
&&\quad\times\Big\{n(s)-n(s+t)\Big\}\nonumber\\
&&-\int_0^t\d{s}w(t-s)r(t-s)\Big\{r(s)n(s)\nonumber\\
&&\qquad-r(s)n(t)+(t-s)r(t)\partial_tn(t)\Big\}\qquad\quad
\label{K-def}\end{eqnarray} and
\begin{equation} U(t)=v^2\int_0^{\infty}\d{s}\big(t+s\big)w(t+s)r(s)\,.
\label{U-def}\end{equation} Furthermore it is appropriate to define
\begin{equation} D(t)=\int_t^{\infty}\d{s}w(s)r(s)-w(t)\int_0^t\d{s}r(s)
\label{D-def}\end{equation} and
\begin{equation} Z(t)=1+\int_0^t\d{s}\,s\,w(s)r(s)\,.
\label{Z-def}\end{equation}

This allows to rewrite the mean field equation \citeq{r-} in the form
\begin{equation} Z(t)\,\partial_tr(t)=J(t)-D(t)r(t)
\label{r-2}\end{equation} and \citeq{n-} as
\begin{equation} Z(t)\,r(t)\partial_t n(t)=U(t)-K(t)\,.
\label{n-2}\end{equation}

\subsection{QFDT- and hierarchical solution}

With \citeq{D-def} and $n(t)\approx0$ for $t<t_p$ the condition
\citeq{condition} means
$D(t_p)=0$.

Depending on $\gamma$ for $v=0$ a QFDT-solution and a hierarchical solution,
respectively, have been proposed \cite{KibachHo93I,KibachHo93II}. These
solutions can easily be obtained from the above equations. Especially the
second scheme requires, however, assumptions which are not fulfilled.
Nevertheless they are presented here for further reference.

\subsubsection{QFDT-solution} In the QFDT-solution $n(t)=n_Q$ for $t>t_p$ is
assumed. From \citeq{condition} one finds
\begin{equation} n_Q+1=-\frac{f'(q_c)}{q_c f''(q_c)}\,.
\label{nQ-def}\end{equation} Without further analysis of the regime $t>t_p$
nothing can be said about the range of validity of this solution or about the
force $F$. This solution was proposed to be valid for $\gamma>1$ and it shows
some similarity to the 1RSB-calculation
\cite{Mez:manifold90,Mez:manifold91}.

\subsubsection{Hierarchical solution} Assume that the integral in
\citeq{r-} for $t>t_p$ is completely determined by its contributions from the
upper and lower bound, respectively. This means that $J(t)\approx 0$. As it
turns out this is the essence of the proposal of the existence of an
ultrametric organization of long time scales. Neglecting the derivative with
respect to $t$ in \citeq{r-2} results in
$D(t)\approx 0$ for $t>t_p$. 

Defining
\begin{equation} m\big(q(t)\big)=\frac{1}{1+n(t)}
\end{equation} Eq.\citeq{D-def} yields
\begin{equation}
\int_q^{\infty}\d{q'}f''(q')m(q')=f''(q)\int_0^q\d{q'}m(q')\,.
\label{m-hierarch}\end{equation} This leads to the differential equation
\begin{equation}
\partial_q \ln m(q)=\frac{f''''(q)}{f'''(q)}-\frac{3f'''(q)}{2f''(q)}
\end{equation} which is solved for $q\ge q_c$ by
\begin{equation}
m(q)=m(q_c)\left\{\frac{1+q}{1+q_c}\right\}^{-\sfrac12(1-\gamma)}
\end{equation} using \citeq{f-def}.

The integration constant $m(q_c)$ is obtained from 
\linebreak
\citeq{m-hierarch} with $q=q_c$
\begin{equation} m(q_c)=-\frac{f'''(q_c)q_c}{2\,f''(q_c)}=R_m
\end{equation} with $R_m$ given in \citeq{R-def}. This is the solution
obtained previously, assuming a hierarchical structure of long time scales
\cite{KibachHo93II}.

\subsection{Evaluation of integrals via effective exponents}

The main problem in a discussion of the solutions of the mean field equations
and also in their numerical integration is in the evaluation of the integrals
listed in the previous section. In the following an approximative scheme is
proposed, which turns out to be very accurate over the whole range of $t$.
This scheme is based on the effective time dependent exponents, which have
been introduced in Section 2.3. It allows to write and evaluate the integrals
in the form
\begin{equation}
\int_0^1\d{x}x^{a-1}(1-x)^{b-1}=\frac{\Gamma(a)\Gamma(b)}{\Gamma(a+b)}\,.
\label{e-integral}\end{equation}

As a first step in this program one introduces "dimen\-sion\-less" quantities
instead of the original ones given in \citeq{J-def} to \citeq{Z-def}:
\begin{equation}
\tilde J(t)=\frac{t\,J(t)}{w(t)\,g^2(t)}\,,
\label{J-.}\end{equation}

\begin{equation}
\tilde K(t)=\frac{t\,K(t)}{k(t)\,w(t)\,g^2(t)}\,,
\label{K-.}\end{equation}

\begin{equation}
\tilde U(t)=\frac{t\,U(t)}{k(t)\,w(t)\,g^2(t)}\,,
\label{U-.}\end{equation}

\begin{equation}
\tilde D(t)=\frac{D(t)}{w(t)\,g(t)}\,,
\label{D-.}\end{equation} and
\begin{equation}
\tilde Z(t)=\frac{Z(t)}{t\,w(t)\,g(t)}\,.
\label{Z-.}\end{equation}

The functions in the integrands are approximated by
\begin{equation} s\,r(s)=g(s)\approx g(t)\left(\frac s t\right)^{\nu(t)}\,,
\label{r(s)-}\end{equation}

\begin{equation} w(s)\approx w(t)\left(\frac s t\right)^{-\alpha(t)}
\label{w(s)-}\end{equation} and
\begin{equation} n(s)\approx n(t)+\frac{k(t)}{\kappa(t)}
\left\{\left(\frac s t\right)^{\kappa(t)}-1\right\}\,.
\end{equation}

This ansatz fulfills \citeq{nu-def}, \citeq{alpha-def},
\citeq{k-def} and \citeq{kappa-def} for $s=t$. The integrals $\tilde J(t)$ and
$\tilde K(t)$ are now of the form proposed above (for the first integral in
\citeq{K-def} a trivial substitution of variables is necessary) and can be
evaluated using
\citeq{e-integral} provided $a>0$ and $b>0$. Otherwise counterterms are
required cancelling the poles of the $\Gamma$-functions. The last three terms
in \citeq{J-def} are of this kind. Depending on the values of the effective
exponents further counterterms might be required.

This yields (in the following the arguments $t$ are dropped)
\begin{eqnarray}
\tilde J&\approx
&\frac{\Gamma(\nu-\alpha)\Gamma(\nu)}{\Gamma(2\nu-\alpha)}\nonumber\\ &&-\frac
1 {\nu}-\frac 1 {\nu-\alpha}+\frac{\nu-1}{\nu+1-\alpha}\qquad\quad
\label{J-.-}\end{eqnarray} and
\begin{eqnarray}
\tilde K&\approx& \frac{1}{\kappa}
\bigg\{\frac{\Gamma(\nu+\kappa)\Gamma(1+\alpha-2\nu-\kappa)}
{\Gamma(1+\alpha-\nu)}\nonumber\\ 
&&\qquad-\frac{\Gamma(\nu)\Gamma(1+\alpha-2\nu-\kappa)}
{\Gamma(1+\alpha-\nu-\kappa)}\nonumber\\
&&\qquad+\frac{\Gamma(\nu-\alpha)\Gamma(\nu)}
{\Gamma(2\nu-\alpha)}-\frac{\Gamma(\nu-\alpha)\Gamma(\nu+\kappa)}
{\Gamma(2\nu+\kappa-\alpha)}\nonumber\\
&&\qquad-\frac{\kappa}{\nu-\alpha+1}\bigg\}\,.
\label{K-.-}\end{eqnarray}

The integral \citeq{U-.} is given for $\nu>0$ and $\alpha-\nu>1$ by
\begin{equation}
\tilde U\approx \frac{v^2\,t^2}{k\,g}\,\frac{\Gamma(\nu)\Gamma(\alpha-\nu-1)}
{\Gamma(\alpha-1)}
\label{U-.-1}\end{equation} otherwise it is approximated by
\begin{equation}
\tilde U\approx \frac{v^2\,t\,Z(\infty)}{k\,w\,g^2}\,.
\label{U-.-2}\end{equation} For $\alpha-\nu>1$ \citeq{Z-.} becomes 
\begin{equation}
\tilde Z\approx \frac{Z(\infty)}{t\,w\,g}
\label{Z-.-2}\end{equation} otherwise
\begin{equation}
\tilde Z\approx \frac 1 {1+\nu-\alpha}\,.
\label{Z-.-1}\end{equation}

The mean field equations \citeq{r-2} and \citeq{n-2} are rewritten as
\begin{equation}
\big\{\nu-1\big\}\,\tilde Z=\tilde J-\tilde D
\label{r-.-}\end{equation} and
\begin{equation}
\tilde Z=\tilde U-\tilde K\,.
\label{n-.-}\end{equation}

The exponent $\alpha$,  \citeq{alpha-def}, is
\begin{equation}
\alpha=-2\,g\,\big\{1+n+u\big\}\frac{f'''(q+v^2t^2)}{f''(q+v^2t^2)}
\label{alpha-.-}\end{equation} with
\begin{equation} u=\frac{v^2\,t^2}{g}\,.
\label{u-def}\end{equation}

Differentiation of \citeq{D-def} yields with \citeq{D-.}
\begin{equation} t\partial_t \tilde D=\big\{\alpha-\nu\big\}\tilde
D-2\big\{1-[1+n+u]\,R \big\}
\label{dD-.-}\end{equation} with
\begin{equation} R=-\frac{f'''(q+v^2t^2)}{f''(q+v^2t^2)}\int_0^t\d{s}r(s)\,.
\label{R-.-}\end{equation}

The complete set of mean field equations now involves 15 functions of time:
$q$, $r$,
$g$, $n$, $k$, $\nu$, $\kappa$, $\alpha$, $u$, $R$, $\tilde D$, $\tilde J$,
$\tilde K$,
$\tilde U$ and $\tilde Z$. The corresponding 15 equations are: \citeq{g-},
\citeq{dq-},
\citeq{nu-def}, \citeq{k-def}, \citeq{kappa-def}, \citeq{J-.-}, \citeq{K-.-},
(\ref{U-.-1}/\ref{U-.-2}),  (\ref{Z-.-2}/\ref{Z-.-1}), \citeq{r-.-},
\citeq{n-.-},
\citeq{u-def}, \citeq{alpha-.-}, \citeq{dD-.-} and \citeq{R-.-}. 

Some of the variables can easily be eliminated, but even then one is left with
5 differential equations of first order and 2 implicit ordinary equations for
7 of the above functions of time. This certainly looks complicated. On the
other hand, in each of the scaling regimes to be discussed below only a subset
of variables and equations has to be looked at.

\subsection{The plateau regime}

Having discussed the FDT-solution in Section 4.1 already, the investigation of
the plateau regime follows next. This regime is characterized by
$\big|q(t)-q_c\big|\ll q_c$. Further simplifications are due to 
$t^2 v^2\ll q_c$. Consequently $R$ defined in \citeq{R-.-} can be replaced by
the constant $R_m$ given in \citeq{R-def}. Furthermore $\tilde U\approx0$ and
$u\approx0$ can be used. From
\citeq{alpha-def} with \citeq{w-def} one finds $\alpha\approx0$ and then
\citeq{Z-.-1} yields
$\tilde Z\approx\frac 1 {1+\nu}$.

Eqs.\citeq{J-.-}, \citeq{r-.-} and \citeq{Z-.-1} determine
$\nu=\tilde\nu(\tilde D)$ as a function of $\tilde D$, shown in
Fig.\ref{nu(D)}. 

\mfigure{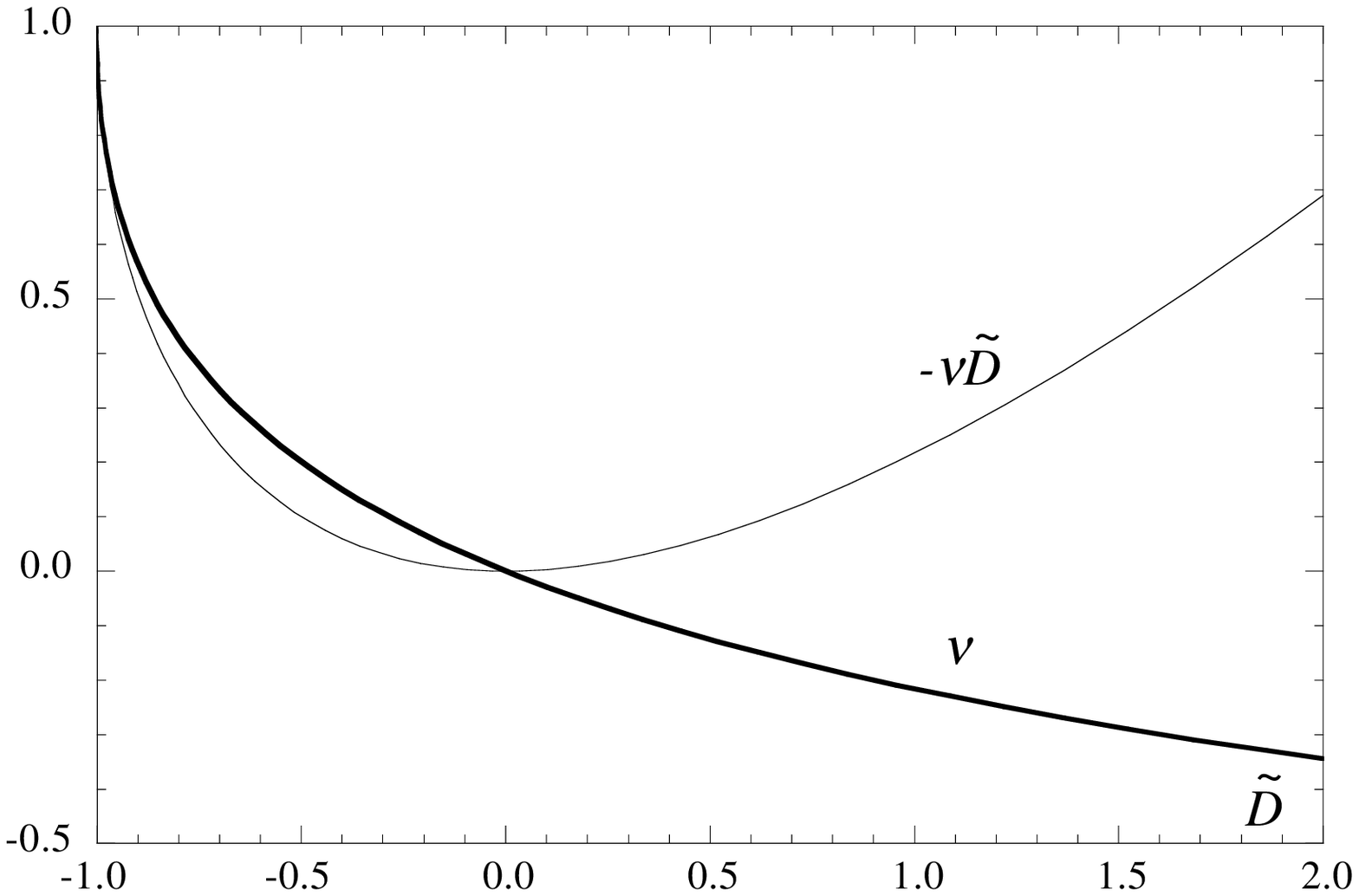} {$\nu(\tilde D)$ and $-\tilde D\nu(\tilde D)$ as functions of
$\tilde D$, see text.}{nu(D)}

Eqs.\citeq{K-.-} and \citeq{n-.-} are solved by
\begin{equation}
\kappa(t)=1-3\nu(t)\,.
\label{ka3nu}\end{equation} Eq. \citeq{dD-.-} now reads
\begin{eqnarray} t\partial_t \tilde D(t)&=&-\tilde\nu\big(\tilde
D(t)\big)\tilde D(t)\nonumber\\ &&-2\big\{1-\big[1+n(t)\big]R_m\Big\}\,.
\label{dD-lin}\end{eqnarray}

The remaining set of equations to be solved consists of \citeq{k-def},
\citeq{kappa-def}, \citeq{J-.-}, \citeq{Z-.-1}, \citeq{r-.-}, \citeq{ka3nu} and
\citeq{dD-lin}. Time enters only via derivatives of the form $t\partial_t$ and
therefore a family of scale invariant solutions exists. This means that the
whole family can be written as a set of functions depending only on $t/t_p$
with arbitrary $t_p$. This is in accordance with the scaling form proposed in
\citeq{scf-pl}.

The time scale $t_p\gg t_0$ can now be fixed such that $\tilde D(t_p)=0$ and
as a consequence $\nu(t_p)=0$ and $\kappa(t_p)=1$. Integrating \citeq{dD-lin}
down to 
$t\ll t_p$ a stable fixpoint $\tilde D(t)\to \tilde D_0>0$ is reached with
\begin{equation} -\tilde\nu(\tilde D_0)\tilde D_0=2\big\{1-R_m\big\}
\end{equation} and $n(t)\to 0$. Eq.\citeq{nuo-} yields $\nu(\tilde
D_0)=\nu_0$. 

For $t_0\ll t \ll t_p$ the function $q(t)$ can be matched to the asymptotic
FDT-solution \citeq{q-FDT} and
\begin{equation}
\hat a(v)=\big(t_p/t_0\big)^{\nu_0}
\label{ahat}\end{equation} results for the scale factor in \citeq{scf-pl}
where $t_p=t_p(v)$ has to be determined later by matching to the asymptotic
regime. It can be shown that the above fixing of
$t_p$ is equivalent to the original definition $q(t_p)=q_c$ within a factor
close to
$1$.

The discussion of the asymptotes for $t\gg t_p$, still within 
$\big|q(t)-q_c\big|\ll q_c$, is more involved. Neglecting the derivative with
respect to time in \citeq{dD-lin} one gets
\begin{equation} -\tilde D\,\tilde\nu(\tilde D)=2\big\{1-(1+n)R_m\big\}
\end{equation} which now determines $\tilde D=\tilde D_1(n)$, 
$\nu_1=\tilde\nu\big(\tilde D_1(n)\big)$ and
$\kappa_1=\tilde\kappa\big(\tilde D_1(n)\big)=1-3\tilde\nu\big(\tilde
D_1(n)\big)$. 

There is a special value $n_x$ such that $\tilde\kappa\big(\tilde
D_1(n_x)\big)=0$ with $\tilde D_1(n_x)=\tilde D_x\approx-0.700084$. This value
is found from
\citeq{dD-lin} as 
\begin{equation} n_x=\frac{1+\frac 1 6 \tilde D_x}{R_m}-1\,.
\label{nx-def}\end{equation} Expanding
\begin{equation}
\tilde\kappa\big(\tilde D_1(n)\big)\approx \big(n-n_x)\,\kappa'
\end{equation} with $\kappa'>0$, Eqs.\citeq{k-def} and \citeq{kappa-def} can
be integrated in closed form. The real solutions are
\begin{equation} n(t)=n_x-\frac{2\,c}{\kappa'}\,\coth\big(c\,\ln(t/t_x)\big)
\label{n1-sr}\end{equation} and
\begin{equation} n(t)=n_x+\frac{2\,c}{\kappa'}\,\tan\big(c\,\ln(t/t_x)\big)
\label{n1-lr}\end{equation} with constants of integration $c$ and $t_x$.

There are now two characteristic values for $n(t)$, the value $n_Q$ defined in
\citeq{nQ-def} and $n_x$ defined above in \citeq{nx-def}. Both are functions
of 
$T$ and $\gamma$. 

The relative magnitude of $n_Q$ and $n_x$ determines which of the above
solutions has to be selected, as will be discussed in the following. This, on
the other hand, determines ultimately whether creep or pinning is observed and
therefore the different phases are characterized by their values of $n_Q$ and
$n_x$. Phase $\cal A$ is defined as the region where $n_x>n_Q>0$, phase $\cal
B$ has $n_Q>n_x>0$ and $n_Q>0>n_x$ holds in phase $\cal C$. The ergodic phase
$\cal O$ with finite friction has already been discussed in Section 4.1

\subsubsection{Phase $\cal A$}

The numerical results shown in Fig.\ref{n-sr} indicate a plateau of $n(t)$
extending to times beyond the upper boundary of the $q$-plateau. This suggests
that the solution
\citeq{n1-sr} has to be used for $t>t_p$. The choice $t_x\sim t_p$ yields the
scaling form \citeq{scf-pl}. For $t\gg t_p$ the requirement $n(t)\to n_Q<n_x$ 
determines
\begin{equation}
c=\frac{\kappa'}{2}\,\big(n_x-n_Q\big)\approx - \frac{\kappa_1}{2}
\end{equation} with $\kappa_1=\tilde\kappa\big(\tilde D_1(n_Q)\big)$. The
parameter $n_Q$ is later shown to agree with the value obtained for the
QFDT-solution \citeq{nQ-def}.

For $t\gg t_p$ the above equations yield with $\nu_1=\tilde\nu\big(\tilde
D_1(n_Q)\big)>0$ and $\kappa_1=1-3\nu_1<0$
\begin{eqnarray}
q(t)&\to&q_c+\big(t_p/t_0\big)^{\nu_0}\,\big(t/t_p\big)^{\nu_1}\,\hat
q_1\nonumber\\ n(t)&\to&n_Q-\big(t/t_p\big)^{\kappa_1}\,\hat n_1\,.
\label{qn-A}\end{eqnarray} For $v\to0$ the constants $\hat q_1$ and $\hat n_1$
depend only on $\gamma$ and $T$.

For $t\ll t_p$
\begin{equation} n(t)\to \hat n_0\big(t/t_p\big)^{\kappa_0}
\label{n0-A}\end{equation} with constant $\hat n_0$ and $\kappa_0>0$. Matching
at $t_0\ll t\ll t_p$ yields the scale factor
\begin{equation}
\otop b(v)=\big(t_0/t_p\big)^{\kappa_0}
\label{b0-match}\end{equation} in \citeq{scf-FDT}. The time scale $t_p=t_p(v)$
is still open and has to be determined later by matching to the asymptotic
regime. The same holds for $n_Q$.

\subsubsection{Phase $\cal B$}

The plateau of $n(t)$ in this phase extends only to the upper boundary of the
$q$-plateau, as shown in Fig.\ref{n-lr}. This indicates that solution
\citeq{n1-lr} now has to be used for $t>t_p$. Its range of validity is
restricted to
$-\sfrac12\pi<c\ln(t/t_x)<\sfrac12\pi$ or 
$\E^{-\pi/2c}t_x<t<\E^{\pi/2c}t_x$. Later it is shown that this range
increases for
$v\to0$ and therefore $c=c(v)\to0$ for $v\to0$. For $t\to t_p$ the solution
should not depend on
$c$, which is the case for the choice
\begin{equation} t_x=\E^{\pi/2c}\,t_p\,.
\label{tx-B}\end{equation} For $t_p\ll t\ll \E^{\pi/c}\,t_p$ the second term
in \citeq{n1-lr} is a small correction and the proposed scaling form
\citeq{scf-pl} and \citeq{n-x} holds over the whole range $t_0\ll t\ll
\E^{\pi/c}\,t_p$.

With $\kappa_1=\tilde\kappa\big(\tilde D_1(n_x)\big)=0$ and
$\nu_1=\tilde\nu\big(\tilde D_1(n_x)\big)=\frac13$, \ $q(t)$ has again the
asymptotic form \citeq{qn-A} for $t\gg t_p$ and the scale factor $\otop b(v)$
is given by \citeq{b0-match}. In this phase the quantities to be determined
later by matching to the asymptotic regime are $t_p=t_p(v)$ and $t_x=t_x(v)$.

\subsubsection{Phase $\cal C$}

In this phase with $n_x<0$ again solution \citeq{n1-sr} has to be used. Since
now $\kappa_1=\tilde\kappa\big(\tilde D_1(n_x)\big)>0$ and $n(t_p)\ll1$, the
appropriate choice of the parameters is
\begin{equation}
\frac{2c}{\kappa'}= - n_x\,;\qquad c=\sfrac12\kappa_1\,.
\end{equation} This yields for $t_x\gg t_p$ and $t_p\ll t\ll t_x$
\begin{equation} n(t)\approx -2 \, n_x\,\big(t/t_x\big)^{\kappa_1}\ll1
\label{n1-C}\end{equation} which means that \citeq{scf-pl} is fulfilled with
$n(t)\approx0$.

The asymptotic form of $q(t)$ is again \citeq{qn-A} with 
$\nu_1=\tilde\nu\big(\tilde D_1(0)\big)$. Again the quantities to be determined
later  by matching to the asymptotic regime are $t_p=t_p(v)$ and $t_x=t_x(v)$.

\subsection{The asymptotic regime}

The discussion so far did not depend on the actual choice of the drift
velocity $v$. On the other hand, there are parameters not determined yet.
These are the time scale $t_p(v)$ for all phases and the second time scale
$t_x(v)$ for phase $\cal B$ and
$\cal C$. For phase $\cal A$ it has to be shown that $n_Q$ actually agrees
with the value obtained in the QFDT-solution, \citeq{nQ-def}. 

The velocity $v$ enters the full set of mean field equations in $\tilde U(t)$, 
(\ref{U-.-1}/\ref{U-.-2}), at various places in the argument of
$f\big(q(t)+v^2t^2\big)$ and its derivatives, and in the definition of $u(t)$, 
\citeq{u-def}. In order to obtain the scaling form proposed in \citeq{scf-as}
for
$t\sim t_a$ it is necessary that $\tilde U(t)$ can be written as
$\tilde U(t)=\bar U(t/t_a)$, which is the case for
\begin{equation}
\frac{v^2\,t_a^2(v)}{\bar b(v)}=1
\label{U-match}\end{equation} assuming the proposed scaling form in
(\ref{U-.-1}/\ref{U-.-2}).

For $t\ll t_a$ the solution in the asymptotic regime has to match the solution
valid in the plateau regime obtained in the preceding section for $t\gg t_p$.
This means
\begin{equation} q(t)\toto_{t\to t_p} q_c+\big(t/t_a\big)^{\nu_1}\,\bar q_0
\end{equation} with constant $\bar q_0$. Comparison with \citeq{qn-A} yields
\begin{equation}
\big(t_p/t_0\big)^{\nu_0}\,t_p^{-\nu_1}\sim t_a^{-\nu_1}\,;\qquad t_p\sim
t_0^{1-\zeta}\,t_a^\zeta
\label{ta-tp}\end{equation} with
\begin{equation}
\zeta=\frac{\nu_1}{\nu_1-\nu_0}\,.
\label{zeta}\end{equation}

The discussion of $n(t)$ has to be done for the different phases separately.

\subsubsection{Phase $\cal A$}

Following \citeq{scf-as} one can write
\begin{equation} n(t)\to n_Q-\big(t/t_a\big)^{\kappa_1}\,\bar b\,\bar n_0
\label{n-as}\end{equation} with constant $\bar n_0$ and $\bar b=\bar b(v)$.
Comparison with \citeq{qn-A} results in
\begin{equation}
\bar b\sim \big(t_a/t_p)^{\kappa_1}\,.
\label{bbar-A}\end{equation} Using this in \citeq{U-match} one finds with
\citeq{ta-tp}
\begin{eqnarray} t_a&\sim&v^{-1+\eta}\,t_0^{\eta}\nonumber\\
t_p&\sim&v^{-(1-\eta)\zeta}\,t_0^{1-(1-\eta)\zeta}
\label{tv-A}\end{eqnarray} with
\begin{equation}
\eta=\frac{\nu_0\,\kappa_1}{2(\nu_1-\nu_0)+\nu_0\kappa_1}\,.
\label{eta-def}\end{equation} Since $\nu_1>0$, $\nu_0<0$ and $\kappa_1<0$ one
finds $\eta>0$.

For $t\sim t_a$ the remaining $v$-dependence in the mean field equations is
not of relevance, since 
\begin{equation} v^2t^2=(v\,t_0)^{2\eta}(t/t_a)^2\ll \bar q(t/t_a)
\end{equation} and 
\begin{equation} u(t)=(v\,t_0)^{2\eta}(t/t_a)^2/\bar g(t/t_a)\ll 1
\end{equation} with $u(t)$ defined in
\citeq{u-def} and  $\bar g(t/t_a)=g(t)$.

For $t\gg t_a$ one obtains $\nu(t)\to1$, $ r(t)\to \bar r_1/t_a$ and 
$q(t)\to \bar q_1\,t/t_a$ where $\bar r_1$ and $\bar q_1$ are constants. This
follows immediately from \citeq{Z-.-2} and \citeq{r-.-} realizing that $\tilde
Z(t)$ diverges $\sim t^{\alpha-2}$, whereas the right hand side of
\citeq{r-.-} remains finite for $t\gg t_a$. This defines an even longer time
scale
\begin{equation} t_a'\sim v^{-1-\eta}\,t_0^{-\eta}\,
\end{equation} where $q(t_a')=(vt_a')^2$. 

For $t_a\ll t\ll t_a'$ Eq. \citeq{alpha-def} yields $\alpha\to \gamma+1$.
Investigating Eq.\citeq{n-.-} one observes that $\tilde Z(t)$, \citeq{Z-.-2},
diverges $\sim t^{\alpha-2}$ and $\tilde U(t)$, \citeq{U-.-1}, $\sim
t^{1-\kappa}$, whereas $\tilde K(t)$, \citeq{K-.-} remains finite. This means
$\kappa(t)\approx 3-\alpha(t)$ and $\kappa(t)\to \kappa_2=2-\gamma$. For
$\gamma<2$ the exponent $\kappa(t)$ increases with increasing $t$ reaching a
value $\kappa_2>0$ and therefore $n(t)$ also starts to increase again for
$t_a\ll t\ll t_a'$ reaching a new constant value $n_2(v)$ for $t\gg t_a'$.
Matching with \citeq{n-as} and
\citeq{bbar-A} yields
\begin{equation} n(t)=n_Q+\big(v\,t_0\big)^{2\eta(1-\kappa_2)}\,\bar n'(t/t_a')
\end{equation} and therefore $\displaystyle n_2(v)-n_Q\sim
\big(v\,t_0\big)^{2\eta(1-\kappa_2)}\toto_{v\to0}0$. For $\gamma>2$ this second
plateau of $n(t)$ is missing, the conclusions are, however, unchanged.

In order to give an estimate of the driving force one has to investigate the
behavior of $\alpha(t)$, Eq.\citeq{alpha-def}, first. In the plateau region
$\alpha(t_p)\approx 0$ was found. Around $t\sim t_a$ it starts to increase and
reaches a value $\alpha(t)\approx\gamma+1$ for $t_a \ll t \ll t_a'$. Around
$t\sim t_a'$ it starts to increase again reaching its asymptotic value
$\alpha(t)=2(\gamma+1)$ for $t\gg t_a'$. The integral in the expression
\citeq{F-} for the driving force gets its main contribution from $s$ such that
$1+\nu(s)-\alpha(s)=0$. For $\gamma>1$ this is fulfilled for $s\sim t_a$ and
$n(t)$ is well approximated by $n_Q$. This yields the following relation
between mean velocity and driving force:
\begin{equation} F=(v\,t_0)^{\eta}\,\bar F
\end{equation} with
\begin{equation}
\bar F=-\frac{2\beta}{1+n_Q}\int_0^{\infty}\d x\,x\,f''\big(\bar
q(x)\big)\,\partial_x
\bar q(x)
\end{equation} which does not depend on $v$. For given force the velocity
increases slowly with increasing force according to
\begin{equation} v\sim F^{1/\eta}\,.
\label{v-creep}\end{equation} This behavior is a form of creep. 

In the one-dimensional case \cite{leDou95} creep has also been found for
$\sfrac12<\gamma<1$ with
\begin{equation} v\sim \E^{a\,F^{\mu}}
\label{v(F)-1}\end{equation} with $\mu=2(\gamma-1)/(2\gamma-1)$. A power law
dependence of the above form
\citeq{v-creep} is obtained for $\gamma=1$. Note that a different definition of
$\gamma$ is used in \cite{leDou95}. The results quoted here refer to the
present definition \citeq{f-def}.

The parameter $n_Q$ can be determined from the condition \citeq{condition}
which is equivalent to the requirement $D(t_p)=0$. The main contribution to
the integral comes from the region where $\nu(s)-\alpha(s)=0$ which is again
the case for $s\sim t_a$. With $n(s)\approx n_Q$ the result
\citeq{nQ-def} is recovered and $n_Q$ indeed agrees with the value derived
within the QFDT-solution.

It is remarkable that the longest time scale in phase $\cal A$ is not the
external time scale $v^{-1}$ but rather $t'_a\sim v^{-1-\eta}$, which is
longer. The ultimate reason for that is the behavior of $\kappa(t)=\kappa_1<0$
at the border between the plateau and the asymptotic regime. This value also
enters the exponent $\eta$,
\citeq{eta-def}.

\subsubsection{Phase $\cal B$}

This is no longer the case in phase $\cal B$. The choice
\begin{equation} t_a\sim v^{-1}
\label{ta-B}\end{equation} allows to rewrite the complete set of mean field
equations for $t\sim t_a$ in terms of functions of $t/t_a$ only, especially
\begin{equation} n(t)=\bar n(t/t_a)\,.
\label{n-B}\end{equation}

In the plateau region $n(t)$ is given by \citeq{n1-lr} and the dependence on
$c$ drops out for $c\ln(t/t_x)\sim\pi/2$, or with \citeq{tx-B} for $t\sim
\E^{\pi/c}t_p$. Matching with \citeq{n-B} yields
\begin{equation} t_a\sim \E^{\pi/c}\,t_p\,.
\label{ta-tp-c}\end{equation} On the other hand, \citeq{ta-tp} results from
matching $q(t)$ and therefore
\begin{equation} c=\frac{\pi}{(\zeta-1)\ln(t_0v)}\,.
\end{equation} This means $c=c(v)\to0$ for $v\to0$ as proposed earlier. With
\citeq{tx-B} the intermediate time scale $t_x$ is
\begin{equation} t_x\sim t_0^{(1-\zeta)/2}\,v^{-(1+\zeta)/2}\,.
\end{equation}

The force necessary to sustain the velocity $v$ has to be calculated from
\citeq{F-}. The main contribution comes as before from the region where
$1+\nu(s)-\alpha(s)=0$, which is the case for $s\sim t_a$. Using the scaling
form $q(t)=\bar q(t/t_a)$,
\citeq{scf-as}, and
\citeq{n-B} yields
\begin{eqnarray} F&=&F_p\nonumber\\ &=&-2\beta\int_0^{\infty}\d
x\,x\,f''(\bar q(x)+x^2)\frac{\partial_x\bar q(x)} {1+\bar n(x)}\qquad
\label{Fp-BC}\end{eqnarray} This does not depend on $v$ and therefore a finite
pinning force $F_p$ exists, which has to be overcome in order to set the
particle in motion.

\subsubsection{Phase $\cal C$}

In phase $\cal C$ again
\begin{equation} t_a=v^{-1}
\end{equation} has to be chosen and matching to \citeq{n1-C} requires
$t_x=t_a$. Otherwise the same arguments as above (phase $\cal B$) hold and
there is again a finite pinning force given by \citeq{Fp-BC}. The main
difference between phase $\cal B$ and $\cal C$ is the value of $n(t)$ for
$t_p\ll t\ll t_a$, which is finite in phase $\cal B$ and essentially zero in
phase $\cal C$.

It is again of interest to compare this result with the one-dimensional case
\cite{Sinai82,leDou95} where a finite pinning force is found for
$\gamma=\sfrac12$, whereas the pinning force diverges for $\gamma<\sfrac12$.
This divergence is due to the fact that no short distance cutoff in the
correlation of the disorder is used in the one-dimensional calculation. This
is also the reason why the phase boundaries in this case do not depend on
temperature.

\subsection{The phase diagram}

Let me summarize the results regarding the phase diagram and the dependence of
$v(F)$. Depending on temperature $T$ and exponent $\gamma$ several phases have
been found. The above considerations yield the  phase diagram shown in
Fig.\ref{phase-fig} and discussed in more detail below. Different dependencies
of the velocity on the driving force are observed in different phases.
Examples are shown in Fig.\ref{v(F)}.

\mfigure{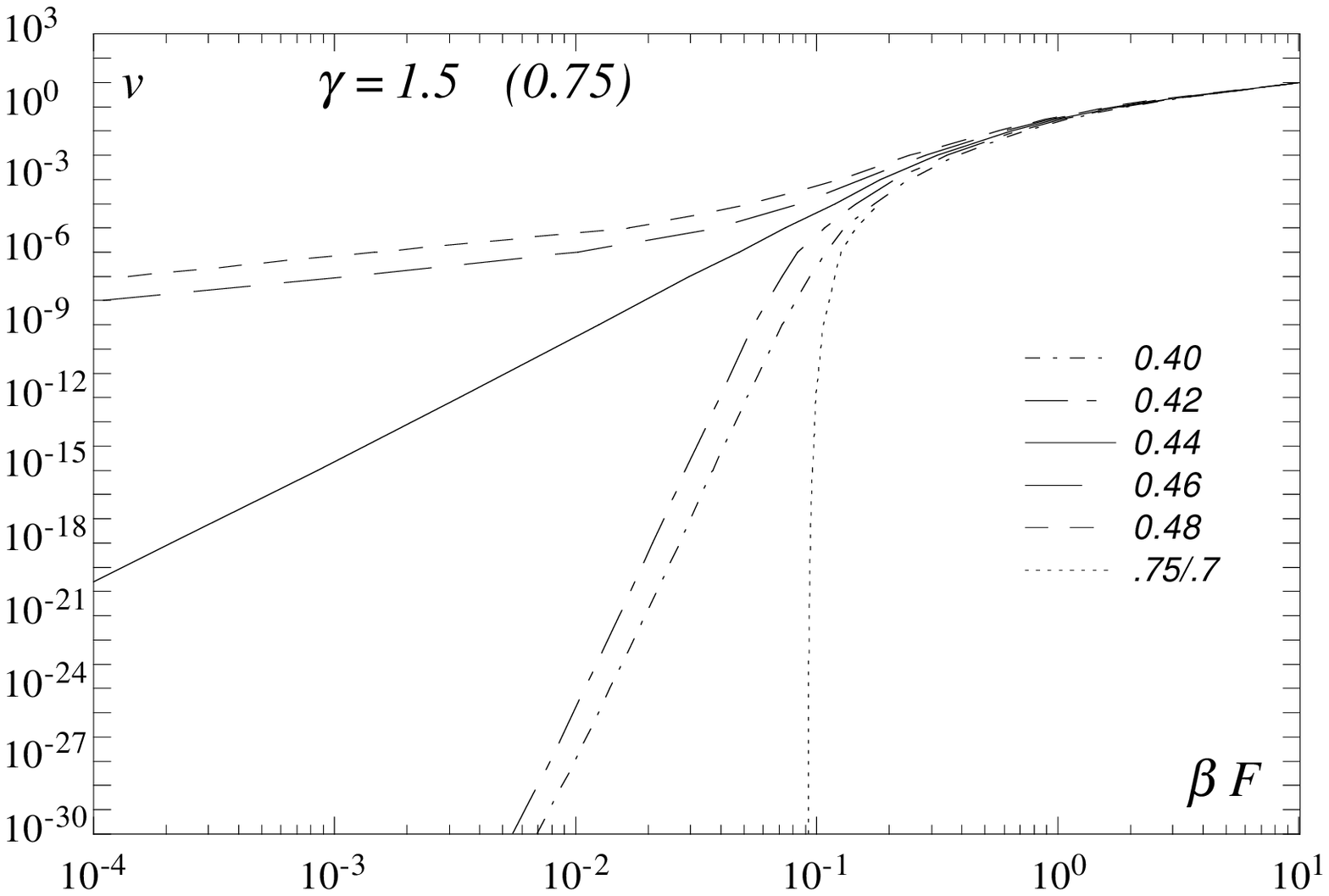} {Drift velocity as a function of the applied force in phase
${\cal O}$, 
${\cal A}$, ${\cal B}$ and at the boundary between phase ${\cal O}$ and ${\cal
A}$.} {v(F)}

The curve marked .75/.7 belongs to $\gamma=0.75$ and $T=0.7$, which is in
phase 
$\cal B$. This point is also used for the examples presented in Section 3.2.
Pinning is clearly observed. With increasing force $v(F)$ becomes linear in
$F$ indicating a finite mobility $m=v/F$. Its value is determined by the first
term in \citeq{F-}, which means that the influence of the random potential
vanishes for high velocities.

The curves marked 0.40 and 0.42 correspond to $\gamma=1.5$ and $T=0.4$ and
$T=0.42$, respectively. Both points are in phase $A$. The temperature $T=0.4$
is the second example in Section 3.2. A power law dependence of $v(F)$ is
found in accordance with the analytic investigations \citeq{v-creep}. For
$\gamma=1.5$ and $T=0.4$ one finds
$\eta=0.053$. With increasing force again the free mobility is found. The curve
marked 0.44 is right at the transition line $\gamma=1.5$, $T=T_c(1.5)=0.4387$,
and
$\eta=0.19$ is found. The remaining curves marked 0.46 and 0.48 are in the high
temperature phase $\cal O$ with $T=0.46$ and $T-0.48$, respectively. For
$v\to0$ a finite mobility is found, which is much smaller than the free value.
It vanishes at the critical temperature $T_c$.

\subsubsection{Drift phase $\cal O$}

For $T>T_c(\gamma)$, \citeq{Tc--}, and $\gamma>1$ a finite mobility $m=v/F$ is
found for $v\to0$. It is given by \citeq{Foo} and it vanishes at the
boundaries of this phase. There are no long time scales.

\subsubsection{Creep phase $\cal A$}

This phase is characterized by $n_Q<n_x$ given in \citeq{nQ-def} and
\citeq{nx-def}. Its boundary with phase $\cal O$ is $T=T_c(\gamma)$ and the
boundary to phase $\cal B$ is determined by $n_Q=n_x$. This yields with
\citeq{R-def},
\citeq{nQ-def} and \citeq{nx-def}
\begin{equation}
\frac{f'(q_c)\,f'''(q_c)}{f''^2(q_c)}=2\,(1+\mbox{$\frac16$}\tilde D_x)\,.
\end{equation} Inserting \citeq{f-def} one finds that this phase boundary is
determined by
$\gamma=\gamma_c$, with 
\begin{equation}
\gamma_c=\frac{1}{1+\frac13\tilde D_x}\approx1.3044\,.
\label{gammac-}\end{equation}

For ${v\to 0}$ one observes creep in the form ${v\sim F^{1/\eta}}$,
\citeq{v-creep}, with $\eta\to0$ for $\gamma\to\gamma_c$.

\subsubsection{Pinning phase $\cal B$ and $\cal C$}

This phase exists for $\gamma<1$ or $T<T_c$ and $\gamma<\gamma_c$. It has a
finite pinning force which has to be overcome in order to set the particle in
motion. Phase $\cal B$ and $\cal C$ are distinct only by their value of $n(t)$
for
$t_p\ll t\ll t_a$. In phase $\cal C$ one finds $n(t)\approx0$ for $ t\ll t_a$
which means that the FDT holds up to this value, whereas it starts to be
violated already for $t\sim t_a$ in phase $\cal A$ and $\cal B$.

\section{Discussion}
\setcounter{equation}{0}

The first aspect of this investigation deals with the motion of a particle in a
correlated random potential with power law decay of the correlations under the
influence of an applied driving force. The phase diagram shows a phase with
finite mobility, a creep phase and  pinning phases. Similar behavior is found
in a  one-dimensional model \cite{Sinai82,leDou95} indicating that the
transitions found are not an artifact of the mean field treatment, which
becomes exact in the limit of infinite dimensionality. In the creep phase
$v(F)$ obeys a power law. Such a behavior is also found in the one-dimensional
case, but only at the boundary between the drift and the creep phase, which is
otherwise ruled by a stretched exponential law. The pinning phase in the
present case has a finite pinning force, which is also found in the
one-dimensional case, but only at the boundary between creep and pinning
phase. Otherwise the pinning force diverges in this calculation, which can be
traced back to the absence of a short distance cutoff of the power law decay
of the correlations.

The numerical results indicate the existence of several scaling regimes which
are   verified by analytic investigations of the asymptotic properties in the
limit of small drift velocity. These regimes are, with increasing time, the
FDT-regime describing a local equilibrium within one of the valleys of the
energy landscape, an intermediate plateau regime where the correlation
function $q(t)$ stays close to the EA-order parameter $q_c$ and where the
characteristic time scale $t_p(v)\sim v^\zeta$, and the asymptotic regime with
characteristic time scale $t_a\sim v^{-1}$. In the creep phase
$\cal A$ this asymptotic regime is ruled by two time scales, $t_a\sim
v^{-1+\eta}$ and 
$t'_a\sim v^{-1-\eta}$. The exponent $\eta$ also determines the power law of
$v(F)$.

The numerical calculations have been performed for a wide range of velocities
including values as small as $10^{-30}$ and over times ranging from $10^{-4}$
to $10^{36}$. This is necessary in order to deduce the full asymptotic
behavior and even at these extreme values part of the structure is not yet
fully developed.

The second aspect relates to glassy non-eqilibrium dynamics of mean field
models. The model investigated here has several advantages in this respect. It
uses Lan\-ge\-vin dynamics which is certainly easier to handle than for
instance Glauber dynamics in Ising type models. Depending on $\gamma$,
continuous as well as discontinuous ergodicity breaking transitions are found.
With applied external force a stationary non-equilibrium state is reached
where correlation and response functions depend only on time differences. The
inverse velocity $v^{-1}$ plays the role of an external long time scale.

The replica treatment of this model \cite{Mez:manifold90,Mez:manifold91}
predicts transitions between a phase with continuous replica symmetry breaking
for
$\gamma<\gamma_c=1$, a 1RSB-phase for $\gamma>1$ and $T<T_{c,\rm 1RSB}$, and a
replica symmetric phase for $\gamma>1$ and $T>T_{c,\rm1RSB}$. The present phase
diagram differs in the sense that $T_c>T_{c,\rm1RSB}$ and $\gamma_c>1$. A
difference in $T_c$ has been observed before in models with discontinuous
transitions
\cite{KibachHo93I,KiTh87,ho:binperc92,CHS93}. This was traced back to the fact
that the states contributing most to the static replica calculation are
different from those relevant for dynamics and are not accessible within
finite time in the thermodynamic limit. The same now appears to be true for
continuous transitions as well.

There have been several proposals regarding the long time dynamics. For the
SK-model Sompolinsky and Zippelius \cite{Somp81,SZ82} proposed a hierarchical
or ultrametric organization of long time scales. This can be rephrased as the
postulate
\cite{Ho:DySK84} that correlation and response functions can be expressed as
functions of $x(t)=1-\ln t/\ln \bar t$ where $\bar t$ is some long external
time scale. It has, however, been shown that this leads to inconsistencies
\cite{Ho:DySK87}. In the present formulation this requirement means $\nu(t)=0$
for times where the hierarchy exists. This is not observed. 

The assumption of an ultrametric hierarchy of time scales leads to results
which are identical to those obtained by Parisi's continuous replica symmetry
breaking scheme
\cite{Par79}. One of the quantities to be compared is the probability of
overlaps
$P(q)$ which is the derivative $P(q)=|\partial_q x(q)|$ of a function $x(q)$
which in dynamics is given by 
\begin{equation} x\big(q(t)\big)=\frac{1}{1+n(t)}\,.
\end{equation} Within this scheme only $x(q)$ is determined, whereas
correlation and response functions are not unique for long times. This is in
contrast to the present investigation where correlation and response functions
are unique for all times.

The function $x(q)$ obtained from replica theory and from the present
investigation are compared in Fig.\ref{x(q)}. The difference between the
results shows again that different states are of relevance in dynamics and
replica theory.

\mfigure{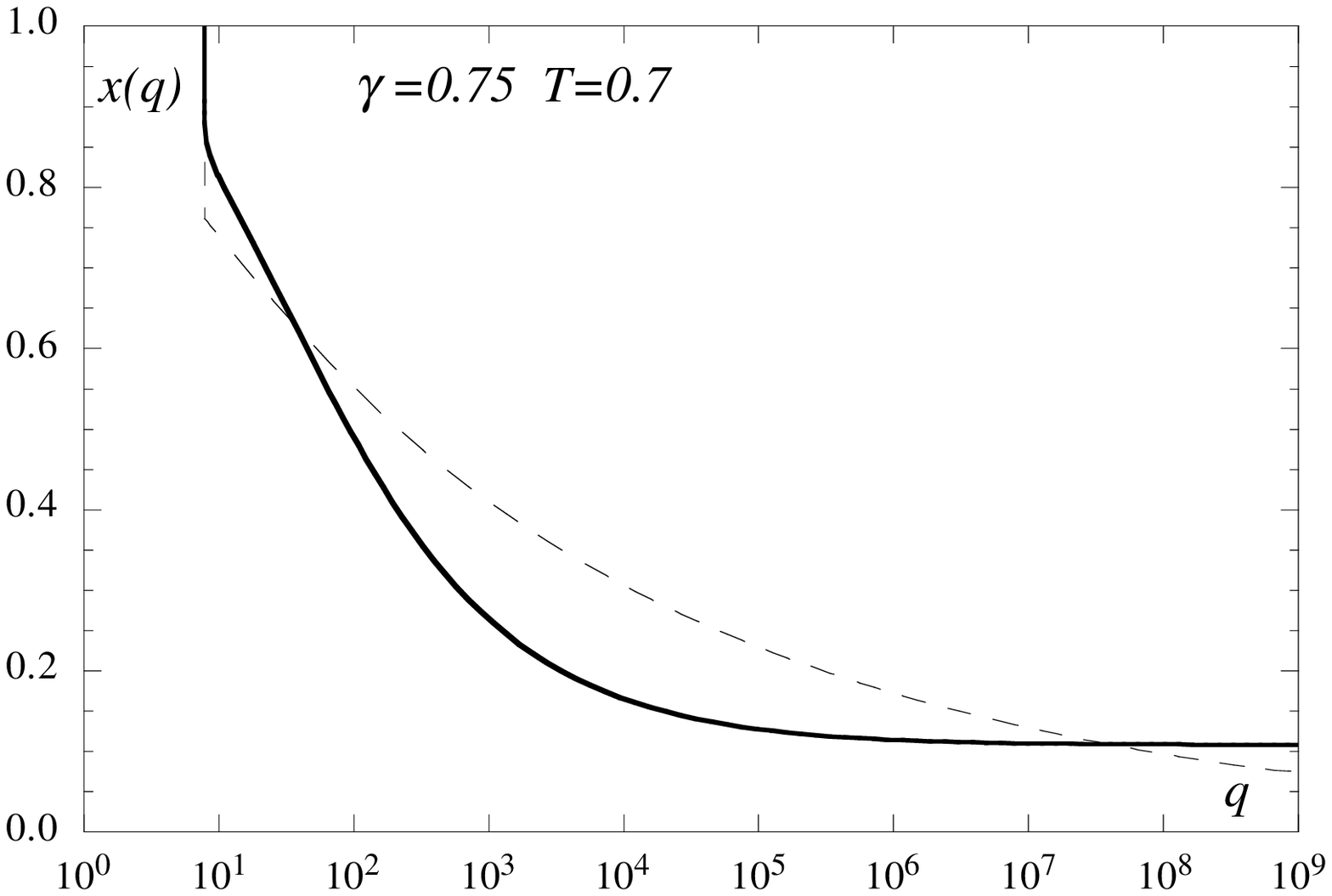} {Function $x(q)$ in phase $\cal B$, see text. The results
obtained in the present investigation (solid) and from replica theory (dashed)
are shown.}{x(q)}

For systems with discontinuous transitions
\cite{KibachHo93I,KiTh87,ho:binperc92,CHS93} the QFDT-solution has been
proposed. It requires $n(t)\to n_Q$ for $t\sim \bar t$. This is actually found
in phase $\cal A$. Within this scheme nothing can be said about the generation
of internal long time scales. Furthermore, the resulting correlation and
response functions are again not uniquely determined, contrary to the present
results.  

There has been considerable interest on mean field dynamics of spin glasses and
related systems with non-equilibrium initial conditions
\cite{CuKu93,FrMe94a,FrMe94b,CuKu94,BCuKuPa95,CuDe95,CuDou95}. In the work of
Cugliandolo and Dean \cite{CuDe95} an exact solution of the spherical SK-model
is reported. It shows aging phenomena although a replica calculation of this
model does not require replica symmetry breaking. The remaining papers deal
with models which have phases with broken replica symmetry (1-step or
continuous). Various proposals are made for the long time properties of
correlation and response functions depending on two time arguments $t$ and
$t'$.  Again for $t-t'\to\infty$ and
$t'\to\infty$ with finite $(t-t')/t'$ the resulting correlation and response
functions could not be determined uniquely. This is likely to be due to an
incomplete analysis of the plateau regime where $t-t'\sim {t'}^\zeta$ with
$0<\zeta<1$. 

A comparison with the present results certainly has to be taken with care
because the specific non-equili\-bri\-um situation is different. Nevertheless
using $t'$ as external time scale one expects that at least the properties for
times $t-t'\ll t'$ can be compared. This includes the plateau regime. A
careful investigation of this regime and the kind of FDT-violation taking
place there seems necessary in this case, too and this is likely to remove the
arbitrariness in the correlation and response functions found so far.

\appendix
\section{Dynamic mean field equations}
\renewcommand{\theequation}{\Alph{section}.\arabic{equation}}
\setcounter{equation}{0}

Time dependent expectation values of products of the variables $\bm{\varrho}$
and
$\hat{\!\bm{\varrho}}$ can be represented as path integrals
\begin{eqnarray}
\av{{\cal O} (\bm{\varrho},\,\hat{\!\bm{\varrho}})}&=&\int \! {\cal
D}\big\{\bm{\varrho},\,\hat{\!\bm{\varrho}}\big\}\,{\cal O}
(\bm{\varrho},\,\hat{\!\bm{\varrho}})\nonumber\\
&&\hskip10pt\times\exp\Big(\!-\!\sum_{\alpha}\left\{S_{\alpha}^o
+S_{\alpha}^v\right\}\Big)\quad
\end{eqnarray} with
\begin{eqnarray}
S_{\alpha}^o&=&\int\d{t}\bigg\{\hat\varrho_{\alpha}(t)^2+i\hat\varrho_\alpha(t)
\Big[\big(\partial_t+\mu_0\big)\varrho_{\alpha}(t)\nonumber\qquad\quad\\
&&\hskip20pt+\sqrt{N}v\delta_{\alpha 1}-\beta F_{\alpha}(t)\Big]\bigg\}
\end{eqnarray} and
\begin{equation} S_{\alpha}^v=i\,\beta\!\int\d{t}\hat\varrho_{\alpha}(t)
\frac{\partial V\big(\bm{\varrho}(t)\big)}{\partial \varrho_{\alpha}(t)}\,.
\end{equation}

Averaging over the disorder with \citeq{VV-av} results in 
\begin{equation}
\sum_{\alpha}S_{\alpha}^v\to \bar
S_v=-\sfrac12\sum_{\alpha\beta}\overline{S_\alpha^vS_{\beta}^v\big.}
\end{equation} and the averaged action is
\begin{eqnarray}
\bar S_v&=&\beta^2\!\int\d{t}\,\d{t'}\bigg\{f'\big(x(t,t')\big)
\sum_{\alpha}\hat{\varrho}_{\alpha}(t)\hat{\varrho}_{\alpha}(t')\nonumber\\
&&+2f''\big(x(t,t')\big)\frac1 N\sum_{\alpha\beta}\hat{\varrho}_{\alpha}(t)
\hat{\varrho}_{\beta}(t')\nonumber\\
&&\times\Big[\varrho_{\alpha}(t)-\varrho_{\alpha}(t')
+\sqrt{N}v(t-t')\delta_{\alpha 1}\Big]\quad\nonumber\\
&&\times\Big[\varrho_{\beta}(t)-
\varrho_{\beta}(t')+\sqrt{N}v(t-t')\delta_{\beta 1}\Big]\bigg\}\qquad\quad
\end{eqnarray} with
\begin{eqnarray} x(t,t')&=&\frac1N\sum_{\alpha}\big[\varrho_{\alpha}(t)
-\varrho_{\alpha}(t')\big]^2\nonumber\\ &&+\frac {2v(t-t')}
{\sqrt{N}}\big[\varrho_1(t)-\varrho_1(t')\big]
\nonumber\\ &&+v^2\big[t-t'\big]^2\,.
\end{eqnarray} This allows to calculate correlation functions \citeq{q-def},
response functions
\citeq{r-def}, using
\begin{equation} r(t-t')=\frac 1 N\sum_{\alpha}
\av{\varrho(t)i\hat{\varrho}(t')}\,,
\end{equation} and other expectation values like $W(t)$, given by
\citeq{VV-av} and \citeq{W-def}.

In mean field theory, see e.g. \cite{KibachHo93I}, an effective action is
introduced replacing appropriate terms in the averaged action by their
expectation values such that the whole problem separates. This effective
action is in the present case
\begin{eqnarray}
S^{eff}_{\alpha}&=&S^o_{\alpha}+\sfrac12\!\int\d{t}\,\d{t'}W(t-t')
\hat{\varrho}_{\alpha}(t)\hat{\varrho}_{\alpha}(t')\nonumber\\
&&+\,i\!\int\d{t}\!\!\int^t\d{t'}w(t-t')r(t-t')
\hat{\varrho}_{\alpha}(t)\nonumber\\
&&\times\big[\varrho_{\alpha}(t)-\varrho_{\alpha}(t')
+\sqrt{N}v(t-t')\delta_{\alpha 1}\big]\,.\qquad\;\;
\end{eqnarray} The identity
\begin{equation}
\av{{\cal O}(\varrho,\hat{\varrho})\frac{\delta S(\varrho,\hat{\varrho})}
{\delta \hat{\varrho}(t)}} =\av{\frac{\delta {\cal
O}(\varrho,\hat{\varrho})}{\delta \hat{\varrho}(t)}}
\end{equation} can be used to derive the equations of motion and the driving
force. The choice
${\cal O}(\varrho,\hat{\varrho})=\hat{\varrho}(0)$ and the requirement of
causality
$r(t)=0$ for ${t\le0}$ yields for $\alpha>1$ the first mean field equation
\citeq{r-}. With ${\cal O}(\varrho,\hat{\varrho})=\varrho(0)$ for $\alpha>1$
and \citeq{q-def} the second mean field equation \citeq{q-} is found. The
driving force \citeq{F-} is obtained from ${\cal O}(\varrho,\hat{\varrho})=1$
for $\alpha=1$.

\setcounter{section}{-1}

\begin{acknowledgement} I want to thank L. Cugliandolo, P.L. le Dousal, S.
Franz, H. Kinzelbach, R. K\"uhn, J. Kurchan, M. M\'ezard and A. Zippelius for
useful discussions. This work was partially supported by EC-contract
CHRX-CT92-0063.
\end{acknowledgement}



\end{document}